\documentclass[12pt,preprint]{aastex}

\shorttitle{GRB X-ray Flares: Spectra}
\shortauthors{Falcone et al.}

\begin{document}


\title{The First Survey of X-ray Flares from Gamma Ray Bursts Observed by {\it{Swift}}: Spectral Properties and Energetics}


\author{
A.~D.~Falcone\altaffilmark{1,2},
D.~Morris\altaffilmark{1},
J.~Racusin\altaffilmark{1},
G.~Chincarini\altaffilmark{3,4},
A.~Moretti\altaffilmark{3},
P.~Romano\altaffilmark{3,4},
D.~N.~Burrows\altaffilmark{1},
C.~Pagani\altaffilmark{1},
M.~Stroh\altaffilmark{1},
D.~Grupe\altaffilmark{1},
S.~Campana\altaffilmark{3},
S.~Covino\altaffilmark{3},
G.~Tagliaferri\altaffilmark{3},
R.~Willingale\altaffilmark{5},
N.~Gehrels\altaffilmark{6}
}

\altaffiltext{1}{Department of Astronomy \& Astrophysics, Pennsylvania State University, University Park, PA 16802, USA}
\altaffiltext{2}{corresponding author email: afalcone@astro.psu.edu}
\altaffiltext{3}{INAF -- Osservatorio Astronomico di Brera, Merate, Italy}
\altaffiltext{4}{Universit\`a degli studi di Milano-Bicocca, Dipartimento di Fisica, Milano, Italy}
\altaffiltext{5}{Department of Physics \& Astronomy, University of Leicester, Leicester LE1 7RH, UK}
\altaffiltext{6}{NASA/Goddard Space Flight Center, Greenbelt, MD}


\begin{abstract}

Observations of gamma ray bursts (GRBs) with Swift produced the
initially surprising result that many bursts have large X-ray flares
superimposed on the underlying afterglow.  The flares were sometimes
intense, had rapid rise and decay phases, and occurred late relative
to the ``prompt'' phase.  One remarkable flare was observed by XRT with a flux $\ge$500$\times$ the
afterglow and a fluence comparable to the prompt GRB fluence.  Many GRBs have several flares,
which are sometimes distinct and sometimes overlapping.  Short,
intense, repetitive, and late flaring can be most easily understood
within the context of the standard fireball model with the internal
engine that powers the prompt GRB emission in an active state at late
times.  However, other models for flares have been proposed.  The
origin of the flares can be investigated by comparing the spectra
during the flares to those of the afterglow and the initial prompt
emission.  In this work, we have analyzed all significant X-ray flares
from the first 110 GRBs observed by Swift.  From this sample 33 GRBs were found to have significant X-ray flares, with 77 flares that were detected above the 3$\sigma$ level.  In addition to temporal analysis presented in a companion paper, a variety of spectral models have been fit to each
flare.  A portion of the X-ray flares had enough counts to allow
complex spectral models, such as Band functions, to be fit to the
data.  In some cases, we find that the spectral fits favor a Band
function model, which is more akin to the prompt emission than to that
of an afterglow.  While some flares do release approximately as much
energy as the ``prompt'' GRB emission, we find that the average
fluence of the flares is $2.4\times10^{-7}$ erg cm$^{-2}$ in the 0.2--10 keV energy band, which is approximately a factor of ten below the average prompt GRB emission fluence.  We also find that the peak energy of the observed flares is
typically in the soft X-ray band, as one should expect due to the selection of the sample from X-ray data.  These results, when combined with those presented in the companion paper on temporal properties of flares, supports the hypothesis that most X-ray flares are late-time activity of the internal engine that spawned the initial GRB; not an afterglow related effect.  

\end{abstract}





\section{Introduction}

Since its launch on 2004 November 20, {\it{Swift}} \citep{geh04} has
provided detailed measurements of numerous GRBs and their afterglows
with unprecedented reaction times.  As of 2006 January 24, 110 bursts
were detected by the Burst Alert Telescope (BAT; Barthelmy et
al. 2004).  Approximately 93\% of these were observed by the narrow
field instruments in less than 200 ks, and most of those were detected
within 200 s (typical reaction time was $\sim$100 s, but occasionally
the BAT detected a burst that was observationally constrained).  The
narrow field instruments are the X-ray telescope (XRT; Burrows et
al. 2005) and the Ultraviolet-Optical Telescope (UVOT; Roming et
al. 2005).  By detecting burst afterglows promptly, and with high
sensitivity, the properties of the early afterglow and extended prompt
emission can be studied in detail for the first time.  This also
facilitates studies of the transition between the prompt emission and
the afterglow.  The rapid response of the pointed X-ray Telescope
(XRT) instrument on {\it{Swift}} has led to the discovery that large
X-ray flares are common in GRBs and occur at times well after the
initial prompt emission.  This paper provides the first survey of the 
spectral features of a large sample of these X-ray flares.

While there are still many unknown factors related to the mechanisms that produce GRB emission, the most commonly accepted model is that of a relativistically expanding fireball with associated internal and external shocks \citep{mes97}.  In this model, internal shocks produce the prompt GRB emission. Observationally, this emission typically has a timescale of $\sim20$ s for long bursts and $\sim$0.2 s for short bursts \citep{mee96}.  The expanding fireball then shocks the ambient material to produce a broadband afterglow that decays quickly (typically as ${\sim}t^{-\alpha}$, with $\alpha\sim1.2$ for the nominal afterglow phase).  When the Doppler boosting angle of this decelerating fireball exceeds the opening angle of the jet into which it is expanding, a steepening of the lightcurve (jet break) is predicted \citep{rho99}.  For a descriptions of the theoretical models of GRB emission and associated observational properties, see \citet{mes02}, \citet{zha04}, \citet{piran05}, \citet{woo93}, and \citet{van00}.  For descriptions of the observational properties of the overall X-ray lightcurve, see \citet{zha06}, \citet{nou06}, \citet{obr06}, and \citet{wil07}.

Several authors have suggested reasons to expect continued activity from the internal engine of the GRB after the classical ``prompt'' emission time frame.  \citet{kat97} considered a model in which a magnetized disk around a central black hole could lead to continued energy release in the form of internal shocks.  The parameters of this energy release would depend on the complex configuration of the magnetic field and the magnetic reconnection dynamics, but time periods as long as days for the delayed emission were predicted.  \citet{pro05} have speculated that energy release can be repeatedly stopped and restarted at late times by magnetic flux accumulation and subsequent release.  \citet{per06} have suggested that the flares from both short and long bursts can be explained within the context of evolution and fragmentation of a viscous accretion disk.  For short bursts, in particular, \citet{dai06} have suggested that late flares can be explained by magnetic reconnection events driven by the breakout of magnetic fields from the surface of differentially rotating millisecond pulsars, which resulted from a progenitor compact binary star merger.  \citet{kin05} have speculated that episodic accretion processes could explain continued internal engine activity.  These authors expect that fragmentation and subsequent accretion during the collapse of a rapidly rotating stellar core could explain observations of extended prompt emission.  In general, the dominant model of an expanding fireball with internal/external shocks \citep{mes97} allows for continued prompt emission, provided that the internal engine is capable of continuing the energy injection.

A few observations prior to {\it{Swift}} have included indications of flaring from GRBs after the prompt GRB emission phase.  \citet{wat03} used XMM-Newton to detect line emission from GRB 030227 nearly 20 hours after the prompt burst.  They inferred continued energy injection at this late time, and concluded that a nearly simultaneous supernova and GRB event would require sporadic power output with a luminosity in excess of $\sim5\times10^{46}$ erg s$^{-1}$.  \citet{pir05} used Beppo-SAX to observe two GRBs with relatively small X-ray flares.  The X-ray flare times for GRB 011121 and GRB 011211 were reported as t=240 s and t=600 s, respectively.  The spectral parameters of these two X-ray flares were consistent with afterglow parameters, and these flares were interpreted as the onset of the afterglow \citep{pir05}.  Two other examples of flaring and/or late timescale emission can be found in \citet{int03} and \citet{gal06}.  Although not a detection of late flares from a particular GRB, the work of \citet{con02}, in which an ensemble of GRBs was analyzed, should also be mentioned.  In this study, 400 long GRBs detected by the Burst and Transient Source Experiment (BATSE) were analyzed together in the form of a summed lightcurve above 20 keV.  Significant emission was found at late times (at least to 1000 s).  There are several possible explanations for this emission that do not require flares, but flares at various times are certainly one possible explanation.

\citet{bur05b} provided the initial report that two bursts detected by {\it{Swift}} showed strong X-ray flares.  The first of these, XRF 050406, was an X-ray flash with a short, and relatively weak, X-ray flare that peaked 213 s after $T_{0}$ of the prompt GRB emission.  Due to the fast rise/decay, the most natural explanation for this flare is continued internal engine activity at late times (i.e. delayed prompt emission).  XRF 050406 was analyzed in detail by \citet{romano06}.  Another burst, GRB 050502B, was studied in detail by \citet{fal06} since it was the first dramatic, high-fluence X-ray flare detected.  This flare, which peaked 740 s after the prompt GRB emission, released as much energy in the X-ray band as the prompt GRB released in the 15-150 keV band.  Following these two GRBs with flares, it became clear that this was a common feature of GRBs, as more and more Swift bursts displayed X-ray flares.  Although there are a few interesting cases of optical flares and of flares simultaneous with higher energy emission detected by {\it{Swift}}-BAT, most of these X-ray flares were generally not accompanied by either optical or 15-150 keV emission at a detectable level.  This implied that the peak of the emission was generally in the soft X-ray band.  

There have been several papers studying individual GRBs with X-ray flares \citep{bur05b,fal06,romano06,pag06,cus06,zha06,mor07,goa07,kri07}.  While the detailed study of individual flares is important, it is equally important to look at the properties of the flares in a more general sense to look for general trends and overall mean properties of the flares.  By comparing these overall properties to those of the prompt GRB emission and the afterglow emission, the mechanism of the flare emission may be elucidated.  Furthermore, we can see if there are multiple classes of flares, or if the flare parameters all fall into one uniform distribution.

In this paper and a companion paper \citep{chi07}(referred to as Paper I hereafter), we present the first temporal and spectral study of a statistical sample of X-ray flares within GRBs.  Paper I presents the temporal properties of the sample, and this paper presents the spectral properties of the sample.  The sample includes all bursts, up until 2006 Jan 24, for which {\it{Swift}} detected at least one significant X-ray flare.

\section{The Sample}

The initial sample was chosen by looking at all {\it{Swift}}-XRT light curves, between launch and 2006 January 24, and eliminating the ones that did not show any hint of deviation from a power law decay with typical breaks (see \citet{zha06}, \citet{nou06}, and \citet{obr06} for a discussion of typical lightcurve breaks).  The remaining light curves were then fit with a broken power law decay (referred to as the underlying decay curve and subscripted throughout text with ``{\it{UL}}'') superposed with a power law rise and decay for any flares that appeared above this underlying decay curve.  The start time of the flare was then defined as the time that the power law rise of the flare intersected the underlying decay power law.  Similarly, the stop time of the flare was then defined as the time that the power law decay of the flare intersected the underlying decay power law.  These times, $t_{start}$ and $t_{stop}$, are defined relative to the trigger time, $T_{0}$, of the GRB.  The signal to noise of the flare was then defined, using simple Poisson statistics, as:
\begin{equation}
S/N = \frac{N_{total}-N_{UL}}{\sqrt{N_{total}+N_{UL}}},
\end{equation}
where $N_{total}$ is defined as the total number of photons during the flare time interval, and $N_{UL}$ is defined as the number of photons from the fit to the underlying decay curve during the flare time interval. 
Only the flares with S/N$>3$ were retained in the sample.  This analysis of 110 GRBs resulted in 33 GRBs with at least one significant flare, and it resulted in a total of 77 flare time intervals, which are listed in Table \ref{tbl:flare_sample}.  Some of these 77 time intervals overlap one another so it is not always clear where one flare begins and another ends.  We define the start and end of each flare as described in the analysis section below.  This sample and the sample of Paper I are largely overlapping, but they differ somewhat due to the different approach and goals.  The flares for which temporal properties can be obtained are frequently different from those for which spectral properties can be obtained.

\begin{deluxetable}{cccccc}
\tabletypesize{\scriptsize}
\tablecaption{The Flare Sample \label{tbl:flare_sample}}
\tablewidth{0pt}
\tablehead{\colhead{GRB} & \colhead{Flare} & \colhead{$t_{start}$ (s)} & \colhead{$t_{stop}$ (s)} & \colhead{$t_{peak}$ (s)} & \colhead{S/N}}
\startdata
GRB050219a & 1 & 118 & 453 & 120 & 18.5 \\ 
GRB050406 & 1 & 139 & 361 & 205 & 11.3 \\ 
GRB050421 & 1 & 136 & 165 & 156 & 3.4 \\ 
GRB050502b & 1 & 410 & 1045 & 695 & 145.7 \\ 
GRB050502b & 2 & 19958 & 48591 & 29896 & 7.2 \\ 
GRB050502b & 3 & 50457 & 178280 & 75355 & 18.4 \\ 
GRB050607 & 1 & 94 & 255 & 145 & 10.3 \\ 
GRB050607 & 2 & 255 & 640 & 312 & 25.2 \\ 
GRB050712 & 1 & 88 & 564 & 252 & 31.0 \\ 
GRB050712 & 2 & 302 & 435 & 339 & 12.9 \\ 
GRB050712 & 3 & 415 & 590 & 478 & 8.9 \\ 
GRB050712 & 4 & 788 & 952 & 888 & 3.8 \\ 
GRB050713a & 1 & 101 & 155 & 0 & 11.7 \\ 
GRB050713a & 2 & 155 & 210 & 0 & 3.2 \\ 
GRB050714b & 1 & 285 & 832 & 374 & 19.2 \\ 
GRB050716 & 1 & 155 & 211 & 177 & 11.2 \\ 
GRB050716 & 2 & 315 & 483 & 385 & 13.2 \\ 
GRB050724 & 1 & 78 & 230 & 120 & 102.6 \\ 
GRB050724 & 2 & 63 & 342 & 261 & 33.7 \\ 
GRB050724 & 3 & 13406 & 402320 & 55783 & 19.7 \\ 
GRB050726 & 1 & 151 & 195 & 162 & 3.0 \\ 
GRB050726 & 2 & 219 & 324 & 274 & 12.2 \\ 
GRB050730 & 1 & 210 & 280 & 228 & 20.6 \\ 
GRB050730 & 2 & 323 & 611 & 435 & 51.9 \\ 
GRB050730 & 3 & 611 & 795 & 678 & 33.7 \\ 
GRB050730 & 4 & 9654 & 12578 & 10319 & 33.2 \\ 
GRB050802 & 1 & 312 & 457 & 435 & 3.8 \\ 
GRB050803 & 1 & 513 & 879 & 753 & 5.8 \\ 
GRB050803 & 2 & 889 & 1516 & 1116 & 4.3 \\ 
GRB050803 & 3 & 4455 & 5703 & 5367 & 5.8 \\ 
GRB050803 & 4 & 7345 & 27698 & 22669 & 14.2 \\ 
GRB050803 & 5 & 7646 & 13093 & 11613 & 14.0 \\ 
GRB050803 & 6 & 17240 & 27698 & 18873 & 5.1 \\ 
GRB050814 & 1 & 1133 & 1974 & 1350 & 3.0 \\ 
GRB050814 & 2 & 1633 & 2577 & 2138 & 6.1 \\ 
GRB050819 & 1 & 56 & 253 & 174 & 11.5 \\ 
GRB050819 & 2 & 9094 & 36722 & 19733 & 6.2 \\ 
GRB050820a & 1 & 200 & 382 & 234 & 66.6 \\ 
GRB050822 & 1 & 106 & 190 & 143 & 21.3 \\ 
GRB050822 & 2 & 212 & 276 & 240 & 8.4 \\ 
GRB050822 & 3 & 390 & 758 & 433 & 50.9 \\ 
GRB050904 & 1 & 343 & 570 & 463 & 41.6 \\ 
GRB050904 & 2 & 857 & 1141 & 953 & 3.0 \\ 
GRB050904 & 3 & 1149 & 1343 & 1235 & 4.2 \\ 
GRB050904 & 4 & 5085 & 9001 & 6765 & 23.0 \\ 
GRB050904 & 5 & 16153 & 24866 & 17329 & 22.1 \\ 
GRB050904 & 6 & 18383 & 38613 & 24156 & 19.5 \\ 
GRB050904 & 7 & 25618 & 30978 & 29392 & 21.6 \\ 
GRB050908 & 1 & 129 & 306 & 145 & 7.3 \\ 
GRB050908 & 2 & 339 & 944 & 404 & 14.0 \\ 
GRB050915a & 1 & 55 & 170 & 111 & 14.3 \\ 
GRB050916 & 1 & 16755 & 32357 & 18898 & 20.1 \\ 
GRB050922b & 1 & 357 & 435 & 377 & 12.1 \\ 
GRB050922b & 2 & 476 & 560 & 497 & 5.6 \\ 
GRB050922b & 3 & 630 & 1541 & 827 & 39.4 \\ 
GRB051006 & 1 & 115 & 148 & 132 & 9.6 \\ 
GRB051006 & 2 & 132 & 201 & 162 & 7.5 \\ 
GRB051006 & 3 & 330 & 749 & 495 & 7.5 \\ 
GRB051016b & 1 & 374 & 1940 & 483 & 3.1 \\ 
GRB051117a & 1 & 2 & 4322 & 157 & 117.6 \\ 
GRB051117a & 2 & 134 & 2794 & 380 & 124.1 \\ 
GRB051117a & 3 & 292 & 1313 & 628 & 70.4 \\ 
GRB051117a & 4 & 574 & 2695 & 926 & 78.6 \\ 
GRB051117a & 5 & 642 & 1820 & 1097 & 71.0 \\ 
GRB051117a & 6 & 1237 & 3119 & 1335 & 95.3 \\ 
GRB051117a & 7 & 659 & 3126 & 1535 & 85.8 \\ 
GRB051210 & 1 & 115 & 152 & 132 & 4.4 \\ 
GRB051227 & 1 & 86 & 245 & 120 & 16.7 \\ 
GRB060108 & 1 & 193 & 429 & 285 & 2.1 \\ 
GRB060108 & 2 & 4951 & 37986 & 10471 & 6.1 \\ 
GRB060109 & 1 & 4305 & 6740 & 4810 & 5.0 \\ 
GRB060111a & 1 & 27 & 196 & 110 & 51.5 \\ 
GRB060111a & 2 & 109 & 203 & 171 & 38.6 \\ 
GRB060111a & 3 & 215 & 433 & 312 & 107.4 \\ 
GRB060115 & 1 & 331 & 680 & 406 & 8.7 \\ 
GRB060124 & 1 & 283 & 644 & 574 & 222.6 \\ 
GRB060124 & 2 & 644 & 1007 & 694 & 179.8 \\ 
\enddata
\end{deluxetable}

\section{Analysis}

The data were reduced using the latest HEAsoft tools (version 6.1.0), including {\it{Swift}} software version 2.0, and the latest response (version 8) and ancillary response files (created using xrtmkarf) available in CALDB at the time of analysis.  Data were screened with standard parameters, including the elimination of time periods when the CCD temperature was warmer than -48$^\circ$ C.  When analyzing WT data, only grades 0--2 were included, and when using PC mode data, only grades 0--12 were included.  Source and background regions were both chosen in a way that avoids overlap with serendipitous sources in the image.  For PC mode data, the source region was typically a 20 pixel (47 arcsec) radius circle.  The background region was typically a circle with a radius of 60 pixels chosen in a source-free region (40 pixels if the field was crowded).  All quoted errors are 1$\sigma$ unless otherwise stated.

In order to avoid pile-up effects in some of the higher count rate PC mode data ($>$ 0.5 c/s), an annular source extraction region was used with an inner radius that varied as a function of rate.  WT mode data is free of significant pile-up effects for nearly all of the flares.  However, pile-up does begin to have a marginal systematic effect on WT mode data above ~100 count/s.  For the few flares in this sample with a brief excursion above 100 count/s, the effect of pile-up is insignificant on this analysis since it averages the spectrum over the entire time interval of the flare, and as a result, the vast majority of flare photons are not affected by pile-up.

\subsection{Light Curve Analysis}

The light curves, and the corresponding temporal analysis, are presented in Paper I.  However, it is worth mentioning a few of the temporal analysis issues related to the spectral analysis presented here.  In particular, there are a few differences between the approaches of this paper and Paper I.  The $t_{start}$ and $t_{stop}$ times in our analysis were not constrained to be exactly the same as those of Paper I.  However, they are approximately the same and the differences are irrelevant for the purposes of fitting the spectra.  The small differences arise from the fact that this analysis fits temporal power law curves to the rise and decay portions of the flares, whereas Paper I fits Gaussians to the flares.  The points on the light curve where these power laws intersect the underlying decay curve power law are defined as $t_{start}$ and $t_{stop}$, and they are reported in Table \ref{tbl:flare_sample}.  This method allows us to easily define a temporal region for performing spectral fits on flare data, even if the flare is missing a large fraction of its lightcurve for any reason.  We are still able to fit a spectrum to large flares that are missing some data on the rising or falling portion of the flare, even if they do not have a well constrained temporal fit.  When calculating the fluence in that portion of the flare, a correction factor is applied based on a power law extrapolation of the flare light curve.  To define the underlying decay curve, we use multiply broken power laws that account for the various phases of the GRB and afterglow light curve decays \citep{nou06, zha06, obr06}.

In some cases, the time range for spectral extraction did not include the entire flare time range due to reasons such as incomplete light curves or overlapping flares.  The time regions used for spectral extraction are shown in Table \ref{tbl:table_time}.  This table shows the times for flare spectral extraction and underlying light curve spectral extraction.  In some cases, the underlying spectral extraction used multiple time regions to improve statistics.  In other cases, there was one large contiguous time period for underlying spectral extraction.

\begin{deluxetable}{cccccccccc}
\tabletypesize{\tiny}
\tablecaption{The time regions used for spectral extraction of flares and the time regions used for the extraction of underlying light curve spectra for each flare are tabulated.  In many cases, the underlying spectra was constrained with one time region with sufficient photons to obtain spectral parameters, and thus there are no entries in the last four columns.  In some cases, statistics were maximized by using multiple time regions for the underlying portion.  In a few instances there are dashes for all underlying time regions, indicating that the canonical value for the underlying spectral index was used, as described in the text.  \label{tbl:table_time}}
\tablewidth{0pt}
\tablehead{ & & \colhead{Flares} & & \colhead{Underlying} & & & & & \\ \colhead{GRB} & \colhead{Flare} & \colhead{$t_{begin}$ (s)} & \colhead{$t_{end}$ (s)} & \colhead{$^{(1)}t_{begin}$ (s)} & \colhead{$^{(1)}t_{end}$ (s)} & \colhead{$^{(2)}t_{begin}$ (s)} & \colhead{$^{(2)}t_{end}$ (s)} & \colhead{$^{(3)}t_{begin}$ (s)} & \colhead{$^{(3)}t_{end}$ (s)}}
\startdata
GRB050219a & 1 & 118 & 453 & 670 & 29603 & --- & --- & --- & --- \\ 
GRB050406 & 1 & 139 & 361 & 1447 & 919330 & --- & --- & --- & --- \\ 
GRB050421 & 1 & 136 & 165 & 167 & 488 & --- & --- & --- & --- \\ 
GRB050502b & 1 & 410 & 1045 & 5384 & 20369 & 161890 & 299820 & --- & --- \\ 
GRB050502b & 2 & 19958 & 48591 & 57 & 355 & 1545 & 19958 & 178280 & 264880 \\ 
GRB050502b & 3 & 50457 & 178280 & 57 & 355 & 1545 & 19958 & 178280 & 264880 \\ 
GRB050607 & 1 & 94 & 255 & 92 & 94 & 685 & 20997 & --- & --- \\ 
GRB050607 & 2 & 255 & 640 & 92 & 94 & 685 & 20997 & --- & --- \\ 
GRB050712 & 1 & 88 & 299 & 5157 & 105060 & --- & --- & --- & --- \\ 
GRB050712 & 2 & 302 & 435 & 5151 & 77682 & --- & --- & --- & --- \\ 
GRB050712 & 3 & 415 & 590 & 5074 & 63858 & --- & --- & --- & --- \\ 
GRB050712 & 4 & 788 & 952 & 5074 & 63858 & --- & --- & --- & --- \\ 
GRB050713a & 1 & 101 & 155 & 3541 & 399630 & --- & --- & --- & --- \\ 
GRB050713a & 2 & 155 & 210 & 3541 & 399630 & --- & --- & --- & --- \\ 
GRB050714b & 1 & 285 & 542 & 3639 & 139690 & --- & --- & --- & --- \\ 
GRB050716 & 1 & 155 & 211 & 105 & 155 & 211 & 331 & --- & --- \\ 
GRB050716 & 2 & 315 & 483 & 211 & 331 & --- & --- & --- & --- \\ 
GRB050724 & 1 & 78 & 230 & 433 & 27350 & --- & --- & --- & --- \\ 
GRB050724 & 2 & 222 & 342 & 433 & 27350 & --- & --- & --- & --- \\ 
GRB050724 & 3 & 13406 & 402320 & 433 & 27350 & --- & --- & --- & --- \\ 
GRB050726 & 1 & 151 & 195 & 324 & 12646 & --- & --- & --- & --- \\ 
GRB050726 & 2 & 219 & 324 & 324 & 8358 & --- & --- & --- & --- \\ 
GRB050730 & 1 & 210 & 280 & 132 & 210 & 280 & 313 & --- & --- \\ 
GRB050730 & 2 & 323 & 611 & --- & --- & --- & --- & --- & --- \\ 
GRB050730 & 3 & 611 & 795 & --- & --- & --- & --- & --- & --- \\ 
GRB050730 & 4 & 9654 & 12578 & 4366 & 6863 & 26422 & 99149 & --- & --- \\ 
GRB050802 & 1 & 312 & 457 & 494 & 2873 & --- & --- & --- & --- \\ 
GRB050803 & 1 & 513 & 879 & 34808 & 778510 & --- & --- & --- & --- \\ 
GRB050803 & 2 & 889 & 1516 & 34808 & 778510 & --- & --- & --- & --- \\ 
GRB050803 & 3 & 4455 & 5703 & 34808 & 778510 & --- & --- & --- & --- \\ 
GRB050803 & 4 & 7345 & 27698 & 34808 & 778510 & --- & --- & --- & --- \\ 
GRB050803 & 5 & 10396 & 13093 & 34808 & 778510 & --- & --- & --- & --- \\ 
GRB050803 & 6 & 17240 & 27698 & 34808 & 778510 & --- & --- & --- & --- \\ 
GRB050814 & 1 & 1133 & 1974 & 5646 & 8644 & 32429 & 98328 & --- & --- \\ 
GRB050814 & 2 & 1633 & 2577 & 5774 & 8741 & 32794 & 96149 & --- & --- \\ 
GRB050819 & 1 & 154 & 193 & --- & --- & --- & --- & --- & --- \\ 
GRB050819 & 2 & 9094 & 36722 & 475 & 7975 & 36722 & 55757 & --- & --- \\ 
GRB050820a & 1 & 200 & 258 & 4811 & 5099900 & --- & --- & --- & --- \\ 
GRB050822 & 1 & 106 & 190 & 5692 & 4932900 & --- & --- & --- & --- \\ 
GRB050822 & 2 & 212 & 276 & 5911 & 4795400 & --- & --- & --- & --- \\ 
GRB050822 & 3 & 415 & 616 & 4714 & 5628400 & --- & --- & --- & --- \\ 
GRB050904 & 1 & 343 & 570 & 586 & 868 & --- & --- & --- & --- \\ 
GRB050904 & 2 & 857 & 1141 & 588 & 876 & --- & --- & --- & --- \\ 
GRB050904 & 3 & 1149 & 1343 & 588 & 861 & --- & --- & --- & --- \\ 
GRB050904 & 4 & 5085 & 7110 & 581 & 865 & --- & --- & --- & --- \\ 
GRB050904 & 5 & 16153 & 18205 & 586 & 873 & --- & --- & --- & --- \\ 
GRB050904 & 6 & 22221 & 25379 & 586 & 873 & --- & --- & --- & --- \\ 
GRB050904 & 7 & 27854 & 30978 & 586 & 873 & --- & --- & --- & --- \\ 
GRB050908 & 1 & 129 & 306 & --- & --- & --- & --- & --- & --- \\ 
GRB050908 & 2 & 339 & 944 & --- & --- & --- & --- & --- & --- \\ 
GRB050915a & 1 & 55 & 170 & 170 & 7424 & --- & --- & --- & --- \\ 
GRB050916 & 1 & 16755 & 32357 & 221 & 13085 & --- & --- & --- & --- \\ 
GRB050922b & 1 & 357 & 435 & 348 & 355 & 435 & 476 & 560 & 623 \\ 
GRB050922b & 2 & 476 & 560 & 348 & 355 & 435 & 476 & 560 & 623 \\ 
GRB050922b & 3 & 630 & 1541 & 348 & 355 & 435 & 476 & 560 & 623 \\ 
GRB051006 & 1 & 115 & 148 & --- & --- & --- & --- & --- & --- \\ 
GRB051006 & 2 & 148 & 180 & --- & --- & --- & --- & --- & --- \\ 
GRB051006 & 3 & 330 & 749 & --- & --- & --- & --- & --- & --- \\ 
GRB051016 & 1 & 374 & 1940 & 3778 & 382750 & --- & --- & --- & --- \\ 
GRB051117a & 1 & 113 & 231 & 16046 & 2410600 & --- & --- & --- & --- \\ 
GRB051117a & 2 & 295 & 571 & 16046 & 2410600 & --- & --- & --- & --- \\ 
GRB051117a & 3 & 571 & 729 & 16046 & 2410600 & --- & --- & --- & --- \\ 
GRB051117a & 4 & 817 & 1044 & 16046 & 2410600 & --- & --- & --- & --- \\ 
GRB051117a & 5 & 1044 & 1237 & 16046 & 2410600 & --- & --- & --- & --- \\ 
GRB051117a & 6 & 1237 & 1466 & 16046 & 2410600 & --- & --- & --- & --- \\ 
GRB051117a & 7 & 1466 & 1737 & 16046 & 2410600 & --- & --- & --- & --- \\ 
GRB051210 & 1 & 115 & 152 & 162 & 426 & --- & --- & --- & --- \\ 
GRB051227 & 1 & 86 & 245 & 258 & 20156 & --- & --- & --- & --- \\ 
GRB060108 & 1 & 193 & 429 & --- & --- & --- & --- & --- & --- \\ 
GRB060108 & 2 & 4951 & 37986 & --- & --- & --- & --- & --- & --- \\ 
GRB060109 & 1 & 4305 & 6740 & 8784 & 325220 & --- & --- & --- & --- \\ 
GRB060111a & 1 & 75 & 137 & 2905 & 712320 & --- & --- & --- & --- \\ 
GRB060111a & 2 & 145 & 204 & 2905 & 712320 & --- & --- & --- & --- \\ 
GRB060111a & 3 & 215 & 433 & 2905 & 712320 & --- & --- & --- & --- \\ 
GRB060115 & 1 & 331 & 680 & 117 & 257 & --- & --- & --- & --- \\ 
GRB060124 & 1 & 283 & 644 & 10605 & 14232 & 32067 & 74305 & --- & --- \\ 
GRB060124 & 2 & 644 & 1007 & 10458 & 14432 & 33443 & 71248 & --- & --- \\ 
\enddata
\end{deluxetable}

\subsection{Spectral Analysis}

Spectral models were fit to data using Xspec version 12.2.0 \citep{arn05}.  Spectra were fit in the 0.3 to 10.0 keV energy range.  A systematic error of 3\% was assigned throughout the energy range due to uncertainties in the response of the instrument, particularly below 0.6 keV.  During fitting, all spectra were binned to $\ge$20 photon/bin, and $\chi^{2}$ statistics were used.

This work is attempting to apply four different models to flare spectral data.  However, it is clear that the flare itself is not the only X-ray emission during the time of the flare.  The underlying afterglow of the GRB is usually already in progress at the time of the flare.  In some cases, this is a small fraction of the flux from the flare and would merely add a small systematic effect to the spectral fit.  However, in other cases, the underlying light curve is a large fraction of the observed X-ray emission, and its effect must be taken into account.  We choose to do this by selecting a region of the light curve before and/or after the flare and fitting the spectra in this time region to a simple absorbed power law, which has the following form.
\begin{equation}
f_{UL} = C_{UL}[e^{-N_{H,UL}\sigma(E)}][\frac{E}{keV}]^{-\Gamma_{UL}}
\end{equation}
where $N_{H,UL}$ is the neutral Hydrogen column density with units $atoms/cm^{-2}$, $\sigma(E)$ is the energy dependent photoelectric absorption cross section \citep{mor83}, $\Gamma$ is the spectral photon index, and C is the normalization constant in units of photons cm$^{-2}$ s$^{-1}$ keV$^{-1}$.  If the fit results in a value for the $N_{H,UL}$ that is significantly below the Galactic $N_{H}$, then the fit is recalculated with $N_{H,UL}$ set equal to the Galactic value taken from \citet{dic90}.  In cases for which there are not enough photons that are obviously part of the underlying light curve (i.e. independent of the flare), a canonical value of $\Gamma=2.0$ is chosen, and the $N_{H}$ is simply tied (i.e. forced to be equal) to the Hydrogen column density in the subsequent flare fitting, $N_{H,flare}$. 

The normalization of this spectral power law is then found by using the power law fit to the temporal decay of the underlying light curve before and/or after the flare.  The underlying temporal power law is extrapolated into the time region of the flare, and it is integrated over that time range to obtain the expected counts from the underlying decay during the flare.  This provides the scale factor by which C$_{UL}$ must be normalized.

Once the underlying spectra and flux have been estimated, as described above, these values are frozen and used as an additive part to the four spectral models for the flares.  We then attempt to fit this additive model to the overall flare plus underlying afterglow spectral data.  The four models we apply are: simple power law, exponentially cut off power law, power law plus blackbody, and Band function.  The application of these non-power-law models, with more complex curvature, has been motivated by the similar application of models to prompt GRB emission surveys (e.g. \citet{ban93,ryd04,kan06}).  For all four models, we also apply photoelectric absorption, which is free to vary.  To illustrate the method, the equations of two of the models are shown below.

Simple absorbed Power Law:
\begin{equation}
f_{total} = C_{flare}[e^{-N_{H,flare}\sigma(E)}][\frac{E}{keV}]^{-\Gamma_{flare}}  +  C_{UL}[e^{-N_{H,UL}\sigma(E)}][\frac{E}{keV}]^{-\Gamma_{UL}}
\end{equation}

Absorbed Exponentially Cut off Power Law:
\begin{equation}
f_{total} = C_{flare}[e^{-N_{H,flare}\sigma(E)}][\frac{E}{keV}]^{-\Gamma_{flare}}[e^{-E/E_{0,flare}}]  +  C_{UL}[e^{-N_{H,UL}\sigma(E)}][\frac{E}{keV}]^{-\Gamma_{UL}}
\end{equation}



The other two models used are the thermal blackbody plus power law model and the Band function model, both of which are added to the underlying afterglow power law model in the same way as the exponential cutoff power law model is added to the underlying model in the equations shown above.  The blackbody model and the GRB Band function model \citep{ban93} are described by \citet{arn05}.  In all cases, $C_{UL}$, $N_{H,UL}$, and $\Gamma_{UL}$ are frozen to the values determined using data from a region before and/or after the flare, and all flare parameters (e.g. $C_{flare}$, $N_{H,flare}$, $\Gamma_{flare}$, $E_{0,flare}$, $E_{c,flare}$, $kT_{flare}$) are free to vary during the fitting process.

\section{Spectral Results}

The results from applying the spectral models described above are presented in this section.  In order to maximize the prospects for having reasonably constrained parameters, we selected only the flares for which there were more than 15 degrees of freedom during the fitting of the power law flare model.  From this point forward in the text, we refer to these flares as {\it{Gold}} flares.

\subsection{Overall Spectral Parameters of Underlying Decay}

\begin{figure}
\includegraphics[scale=0.6]{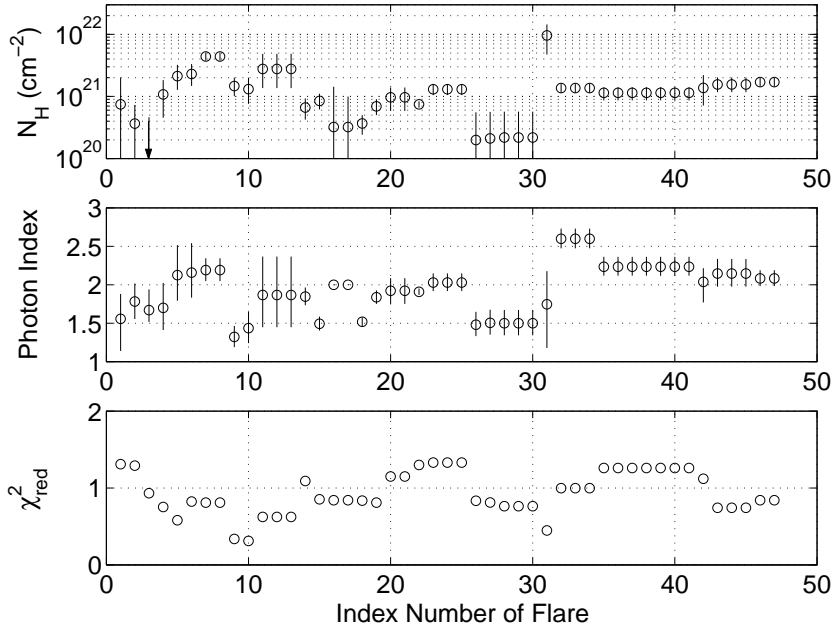}
\caption{Properties of absorbed power law spectral fits to data from a region of the lightcurve in which \emph{no flares} were present, for all GRBs with {\it{Gold}} flares (i.e. these are the spectral parameters of the underlying light curve).  The index number of the flares shown on the x-axis simply refers to the index number for each flare shown in column 1 of Table \ref{tbl:pow_table}.}
\label{fig:properties_underlying_gold}
\end{figure}

The spectral parameters derived by fitting an absorbed power law model to the underlying afterglow of GRBs with {\it{Gold}} flares are shown in Figure \ref{fig:properties_underlying_gold}.  The mean of the photon index distribution is 1.9 with a standard deviation of 0.3.  This is consistent with the typical photon index for GRB afterglows.

\subsection{Overall Spectral Parameters of Flares}

\begin{figure}
\includegraphics[scale=0.8]{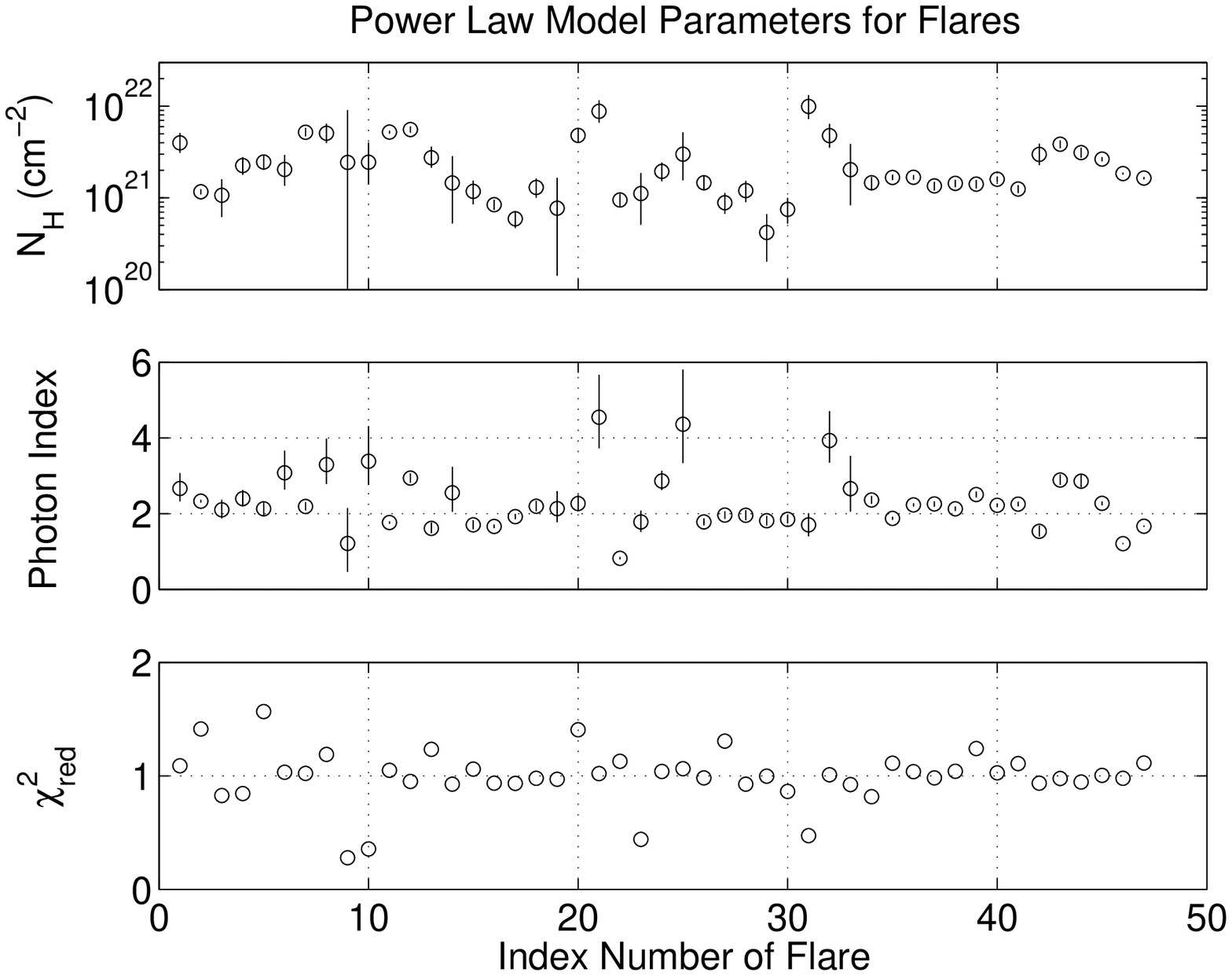}
\caption{Properties of power law spectral fits to flare data for all {\it{Gold}} flares.  The index number of the flares shown on the x-axis simply refers to the index number for each flare shown in column 1 of Table \ref{tbl:pow_table}.  The top panel corresponds to the fit for the neutral Hydrogen column density, the second panel corresponds to the photon index ($\Gamma_{flare}$), and the bottom panel is the reduced $\chi^2$ for each fit.}
\label{fig:properties_pow_gold}
\end{figure}

\begin{figure}
\includegraphics[scale=0.8]{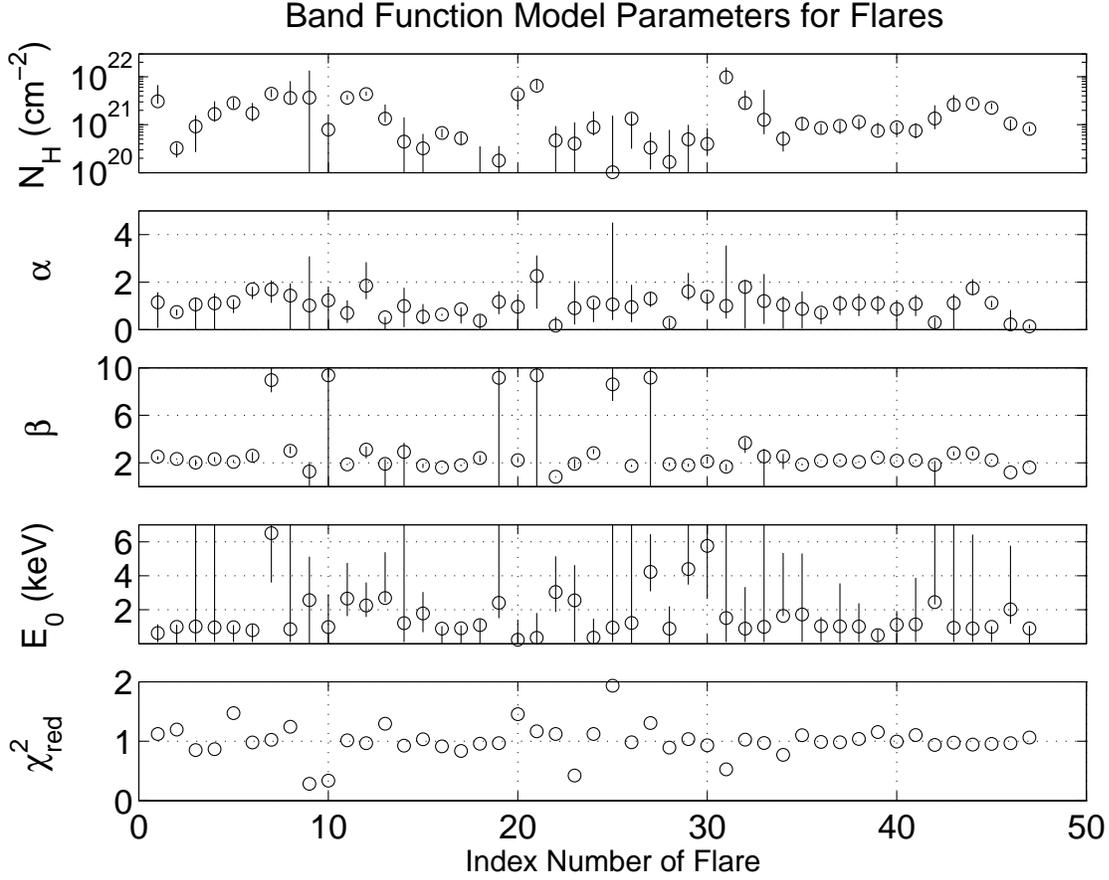}
\caption{Properties of Band function spectral fits to flare data for all {\it{Gold}} flares.  The index number of the flares shown on the x-axis simply refers to the index number for each flare shown in column 1 of Table \ref{tbl:band_table}.  The top panel corresponds to the fit for the neutral Hydrogen column density, the second panel corresponds to the low energy photon index ($\alpha$), and the third panel corresponds to the high energy photon index ($\beta$).  The fourth panel is the e-folding energy ($E_0$), which is related to the peak spectral energy by the relation $E_{peak}=(2+\alpha)E_0$.  The bottom panel is the reduced $\chi^2$ for the fits.}
\label{fig:properties_band_gold}
\end{figure}

\begin{figure}
\includegraphics[scale=0.8]{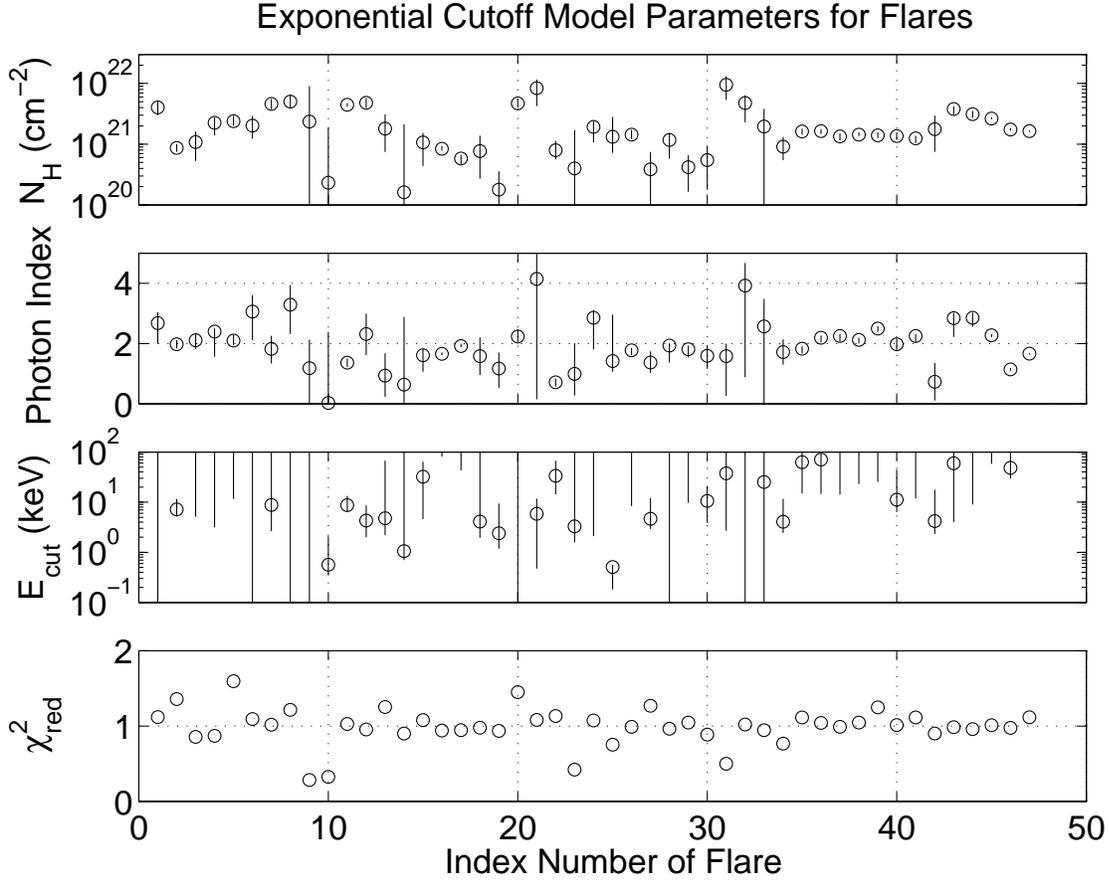}
\caption{Properties of exponentially cutoff power law model spectral fits to flare data for all {\it{Gold}} flares.  The index number of the flares shown on the x-axis simply refers to the index number for each flare shown in column 1 of Table \ref{tbl:cutoff_table}.  Many flares did not provide enough data in this energy band to lead to convergence for the cutoff energy, which is clear from the fact that panel 3 has many data points not shown off the top of the plot (these were set to the 500 keV fitting limit and their lower error bars extend into the plot).}
\label{fig:properties_cutoff_gold}
\end{figure}

\begin{figure}
\includegraphics[scale=0.8]{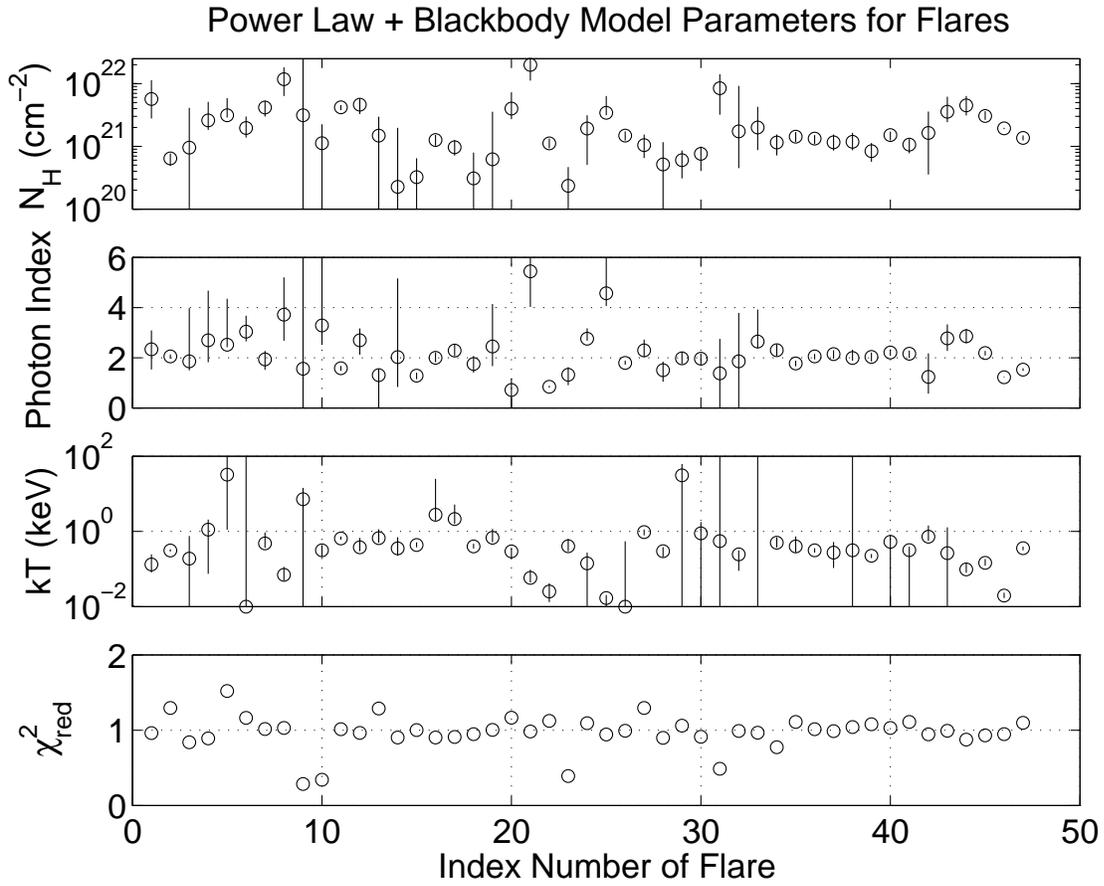}
\caption{Properties of blackbody plus power law model spectral fits to flare data for all {\it{Gold}} flares.  The index number of the flares shown on the x-axis simply refers to the index number for each flare shown in column 1 of Table \ref{tbl:bb_table}.}
\label{fig:properties_bb_pow_gold}
\end{figure}

The spectral parameters derived by fitting the absorbed power law model to the {\it{Gold}} flares are shown in Figure \ref{fig:properties_pow_gold} and Table \ref{tbl:pow_table}.  The index of each flare, shown in column 1 of each table, corresponds to the x-axis of the plots.  A simple absorbed power law can provide a reasonable fit in most cases.  The spectral parameters derived by fitting the absorbed Band function model to the {\it{Gold}} flares are shown in Figure \ref{fig:properties_band_gold} and Table \ref{tbl:band_table}.  Once again, the fit is reasonable in nearly all cases.  The spectral parameters derived by fitting the absorbed exponential cutoff model to the {\it{Gold}} flares are shown in Figure \ref{fig:properties_cutoff_gold} and Table \ref{tbl:cutoff_table}, and the spectral parameters derived by fitting the absorbed power law plus blackbody model to the {\it{Gold}} flares are shown in Figure \ref{fig:properties_bb_pow_gold} and Table \ref{tbl:bb_table}.

\begin{deluxetable}{ccccccc}
\tabletypesize{\tiny}
\tablecaption{Properties of power law spectral fits to {\it{Gold}} flares \label{tbl:pow_table}}
\tablewidth{0pt}
\tablehead{\colhead{Index Number} & \colhead{GRB} & \colhead{Flare} & \colhead{$N_H (10^{20} cm^{-2})$} & \colhead{Photon Index} & \colhead{$\chi^2_{red}$} & \colhead{DOF}}
\startdata
1 & GRB050219 & 1 & 39.9$^{+11.1}_{-9.1}$ & 2.67$^{+0.41}_{-0.34}$ & 1.09 & 38 \\ 
2 & GRB050502 & 1 & 11.7$^{+0.7}_{-0.7}$ & 2.33$^{+0.04}_{-0.04}$ & 1.41 & 328 \\ 
3 & GRB050502 & 3 & 10.7$^{+5.3}_{-4.5}$ & 2.10$^{+0.27}_{-0.23}$ & 0.83 & 31 \\ 
4 & GRB050607 & 2 & 22.5$^{+5.3}_{-4.7}$ & 2.40$^{+0.23}_{-0.20}$ & 0.84 & 33 \\ 
5 & GRB050712 & 1 & 24.6$^{+4.7}_{-4.3}$ & 2.13$^{+0.18}_{-0.17}$ & 1.57 & 57 \\ 
6 & GRB050712 & 2 & 20.5$^{+8.9}_{-6.9}$ & 3.08$^{+0.59}_{-0.44}$ & 1.03 & 18 \\ 
7 & GRB050713 & 1 & 52.1$^{+5.9}_{-5.3}$ & 2.19$^{+0.11}_{-0.11}$ & 1.02 & 188 \\ 
8 & GRB050713 & 2 & 50.7$^{+13.8}_{-10.9}$ & 3.30$^{+0.68}_{-0.51}$ & 1.19 & 48 \\ 
9 & GRB050716 & 1 & 24.4$^{+66.4}_{-24.4}$ & 1.22$^{+0.93}_{-0.75}$ & 0.28 & 56 \\ 
10 & GRB050716 & 2 & 24.5$^{+15.6}_{-10.5}$ & 3.38$^{+0.93}_{-0.63}$ & 0.36 & 55 \\ 
11 & GRB050724 & 1 & 52.1$^{+2.0}_{-1.9}$ & 1.77$^{+0.03}_{-0.03}$ & 1.05 & 330 \\ 
12 & GRB050724 & 2 & 55.6$^{+4.6}_{-4.3}$ & 2.94$^{+0.13}_{-0.12}$ & 0.95 & 54 \\ 
13 & GRB050724 & 3 & 27.4$^{+8.8}_{-6.0}$ & 1.61$^{+0.15}_{-0.13}$ & 1.23 & 22 \\ 
14 & GRB050726 & 2 & 14.6$^{+13.9}_{-9.3}$ & 2.55$^{+0.68}_{-0.50}$ & 0.93 & 37 \\ 
15 & GRB050730 & 1 & 11.8$^{+3.7}_{-3.3}$ & 1.71$^{+0.12}_{-0.12}$ & 1.06 & 58 \\ 
16 & GRB050730 & 2 & 8.4$^{+1.1}_{-1.0}$ & 1.66$^{+0.05}_{-0.05}$ & 0.94 & 187 \\ 
17 & GRB050730 & 3 & 5.9$^{+1.3}_{-1.2}$ & 1.92$^{+0.07}_{-0.07}$ & 0.93 & 106 \\ 
18 & GRB050730 & 4 & 13.0$^{+3.4}_{-3.0}$ & 2.20$^{+0.14}_{-0.13}$ & 0.98 & 81 \\ 
19 & GRB050802 & 1 & 7.7$^{+8.8}_{-6.3}$ & 2.13$^{+0.46}_{-0.36}$ & 0.97 & 30 \\ 
20 & GRB050803 & 5 & 47.9$^{+9.0}_{-7.8}$ & 2.27$^{+0.23}_{-0.21}$ & 1.41 & 34 \\ 
21 & GRB050803 & 6 & 87.8$^{+28.3}_{-21.7}$ & 4.55$^{+1.12}_{-0.82}$ & 1.02 & 18 \\ 
22 & GRB050820 & 1 & 9.5$^{+1.6}_{-1.5}$ & 0.82$^{+0.04}_{-0.04}$ & 1.13 & 202 \\ 
23 & GRB050822 & 1 & 11.1$^{+7.6}_{-6.1}$ & 1.78$^{+0.30}_{-0.27}$ & 0.44 & 27 \\ 
24 & GRB050822 & 2 & 19.4$^{+4.9}_{-4.3}$ & 2.86$^{+0.27}_{-0.24}$ & 1.04 & 31 \\ 
25 & GRB050822 & 3 & 29.9$^{+22.0}_{-14.4}$ & 4.36$^{+1.45}_{-1.03}$ & 1.06 & 18 \\ 
26 & GRB050904 & 1 & 14.6$^{+2.6}_{-2.4}$ & 1.78$^{+0.09}_{-0.09}$ & 0.98 & 182 \\ 
27 & GRB050904 & 4 & 8.9$^{+2.4}_{-2.2}$ & 1.96$^{+0.10}_{-0.10}$ & 1.31 & 38 \\ 
28 & GRB050904 & 5 & 12.0$^{+3.3}_{-3.0}$ & 1.96$^{+0.14}_{-0.13}$ & 0.93 & 26 \\ 
29 & GRB050904 & 6 & 4.2$^{+2.5}_{-2.2}$ & 1.81$^{+0.13}_{-0.12}$ & 1.00 & 22 \\ 
30 & GRB050904 & 7 & 7.5$^{+2.5}_{-2.2}$ & 1.85$^{+0.12}_{-0.11}$ & 0.86 & 24 \\ 
31 & GRB050916 & 1 & 99.4$^{+32.6}_{-26.9}$ & 1.70$^{+0.31}_{-0.30}$ & 0.47 & 20 \\ 
32 & GRB050922 & 1 & 47.9$^{+16.4}_{-12.6}$ & 3.94$^{+0.78}_{-0.58}$ & 1.01 & 99 \\ 
33 & GRB050922 & 2 & 20.3$^{+18.3}_{-12.0}$ & 2.66$^{+0.86}_{-0.60}$ & 0.92 & 45 \\ 
34 & GRB050922 & 3 & 14.6$^{+2.4}_{-2.2}$ & 2.36$^{+0.10}_{-0.10}$ & 0.82 & 116 \\ 
35 & GRB051117 & 1 & 16.7$^{+1.1}_{-1.1}$ & 1.88$^{+0.04}_{-0.04}$ & 1.11 & 342 \\ 
36 & GRB051117 & 2 & 16.8$^{+1.0}_{-0.9}$ & 2.23$^{+0.04}_{-0.04}$ & 1.04 & 318 \\ 
37 & GRB051117 & 3 & 13.5$^{+1.5}_{-1.4}$ & 2.26$^{+0.07}_{-0.07}$ & 0.98 & 181 \\ 
38 & GRB051117 & 4 & 14.4$^{+1.4}_{-1.3}$ & 2.13$^{+0.06}_{-0.06}$ & 1.04 & 226 \\ 
39 & GRB051117 & 5 & 14.1$^{+1.5}_{-1.5}$ & 2.51$^{+0.08}_{-0.08}$ & 1.24 & 184 \\ 
40 & GRB051117 & 6 & 16.0$^{+1.2}_{-1.1}$ & 2.22$^{+0.05}_{-0.05}$ & 1.03 & 265 \\ 
41 & GRB051117 & 7 & 12.5$^{+1.2}_{-1.2}$ & 2.25$^{+0.06}_{-0.06}$ & 1.11 & 223 \\ 
42 & GRB051227 & 1 & 29.9$^{+9.1}_{-7.1}$ & 1.53$^{+0.15}_{-0.14}$ & 0.93 & 24 \\ 
43 & GRB060111 & 1 & 38.5$^{+3.9}_{-3.6}$ & 2.89$^{+0.14}_{-0.13}$ & 0.98 & 118 \\ 
44 & GRB060111 & 2 & 31.1$^{+4.1}_{-3.7}$ & 2.86$^{+0.18}_{-0.17}$ & 0.95 & 76 \\ 
45 & GRB060111 & 3 & 26.5$^{+1.4}_{-1.4}$ & 2.27$^{+0.05}_{-0.05}$ & 1.00 & 297 \\ 
46 & GRB060124 & 1 & 18.4$^{+0.5}_{-0.5}$ & 1.21$^{+0.01}_{-0.01}$ & 0.98 & 681 \\ 
47 & GRB060124 & 2 & 16.4$^{+0.5}_{-0.5}$ & 1.67$^{+0.02}_{-0.02}$ & 1.11 & 536 \\  
\enddata
\end{deluxetable}

\begin{deluxetable}{ccccccccc}
\tabletypesize{\tiny}
\tablecaption{Properties of Band function spectral fits to {\it{Gold}} flares \label{tbl:band_table}}
\tablewidth{0pt}
\tablehead{\colhead{Index Number} & \colhead{GRB} & \colhead{Flare} & \colhead{$N_H (10^{20} cm^{-2})$} & \colhead{$\alpha$} & \colhead{$\beta$} & \colhead{$E_{peak}$ (keV)} & \colhead{$\chi^2_{red}$} & \colhead{DOF}}
\startdata
1 & GRB050219 & 1 & 30.8$^{+36.3}_{-3.3}$ & 1.15$^{+0.43}_{-1.07}$ & 2.52$^{+0.04}_{-0.27}$ & 0.6$^{+0.5}_{-0.5}$ & 1.12 & 36 \\ 
2 & GRB050502 & 1 & 3.3$^{+0.9}_{-1.2}$ & 0.74$^{+0.11}_{-0.14}$ & 2.33$^{+0.04}_{-0.05}$ & 1.0$^{+0.1}_{-1.0}$ & 1.20 & 326 \\ 
3 & GRB050502 & 3 & 9.2$^{+6.5}_{-6.5}$ & 1.06$^{+0.25}_{-1.24}$ & 2.01$^{+0.24}_{-0.26}$ & 1.0$^{+999.0}_{-0.9}$ & 0.85 & 29 \\ 
4 & GRB050607 & 2 & 16.7$^{+13.9}_{-5.2}$ & 1.11$^{+0.41}_{-2.52}$ & 2.31$^{+0.18}_{-0.21}$ & 1.0$^{+66.8}_{-0.9}$ & 0.87 & 31 \\ 
5 & GRB050712 & 1 & 28.2$^{+8.0}_{-6.3}$ & 1.16$^{+0.11}_{-0.46}$ & 2.08$^{+0.16}_{-0.09}$ & 1.0$^{+0.1}_{-0.9}$ & 1.47 & 55 \\ 
6 & GRB050712 & 2 & 17.2$^{+11.3}_{-5.5}$ & 1.70$^{+0.11}_{-0.42}$ & 2.60$^{+0.30}_{-0.45}$ & 0.8$^{+0.3}_{-0.7}$ & 0.98 & 16 \\ 
7 & GRB050713 & 1 & 44.1$^{+10.8}_{-6.7}$ & 1.69$^{+0.38}_{-0.57}$ & 8.97$^{+6.87}_{-1.03}$ & 6.5$^{+21.9}_{-2.9}$ & 1.02 & 186 \\ 
8 & GRB050713 & 2 & 36.1$^{+44.7}_{-8.3}$ & 1.44$^{+0.51}_{-1.58}$ & 3.02$^{+0.34}_{-0.23}$ & 0.9$^{+42.8}_{-0.8}$ & 1.24 & 46 \\ 
9 & GRB050716 & 1 & 36.7$^{+97.4}_{-36.7}$ & 1.02$^{+2.07}_{-8.74}$ & 1.28$^{+0.79}_{-1.28}$ & 2.6$^{+2.6}_{-2.6}$ & 0.28 & 54 \\ 
10 & GRB050716 & 2 & 7.9$^{+7.9}_{-7.4}$ & 1.23$^{+0.59}_{-1.40}$ & 9.37$^{+19.37}_{-9.37}$ & 1.0$^{+1.9}_{-1.0}$ & 0.34 & 53 \\ 
11 & GRB050724 & 1 & 36.5$^{+5.1}_{-3.3}$ & 0.69$^{+0.53}_{-0.42}$ & 1.87$^{+0.06}_{-0.07}$ & 2.7$^{+2.1}_{-1.0}$ & 1.01 & 328 \\ 
12 & GRB050724 & 2 & 43.6$^{+5.5}_{-3.6}$ & 1.85$^{+0.99}_{-0.56}$ & 3.11$^{+0.27}_{-0.74}$ & 2.3$^{+1.4}_{-0.7}$ & 0.97 & 52 \\ 
13 & GRB050724 & 3 & 13.4$^{+13.0}_{-2.1}$ & 0.51$^{+0.02}_{-1.15}$ & 1.91$^{+0.40}_{-1.91}$ & 2.7$^{+2.7}_{-0.2}$ & 1.30 & 20 \\ 
14 & GRB050726 & 2 & 4.4$^{+9.7}_{-4.4}$ & 0.99$^{+0.77}_{-0.88}$ & 2.92$^{+0.77}_{-7.08}$ & 1.2$^{+999}_{-1.1}$ & 0.92 & 35 \\ 
15 & GRB050730 & 1 & 3.2$^{+3.2}_{-3.2}$ & 0.55$^{+0.52}_{-0.33}$ & 1.77$^{+0.16}_{-0.20}$ & 1.8$^{+1.3}_{-1.1}$ & 1.03 & 57 \\ 
16 & GRB050730 & 2 & 6.7$^{+1.5}_{-1.4}$ & 0.63$^{+0.03}_{-0.05}$ & 1.61$^{+0.05}_{-0.03}$ & 0.9$^{+0.1}_{-0.9}$ & 0.91 & 185 \\ 
17 & GRB050730 & 3 & 5.2$^{+1.6}_{-1.4}$ & 0.86$^{+0.07}_{-0.59}$ & 1.80$^{+0.07}_{-0.08}$ & 0.9$^{+0.2}_{-0.9}$ & 0.84 & 104 \\ 
18 & GRB050730 & 4 & 0.9$^{+2.6}_{-0.9}$ & 0.38$^{+0.24}_{-0.96}$ & 2.40$^{+0.24}_{-0.32}$ & 1.1$^{+0.4}_{-1.1}$ & 0.96 & 79 \\ 
19 & GRB050802 & 1 & 1.8$^{+1.8}_{-1.8}$ & 1.17$^{+0.45}_{-0.52}$ & 9.16$^{+19.16}_{-9.16}$ & 2.4$^{+5.6}_{-0.9}$ & 0.97 & 29 \\ 
20 & GRB050803 & 5 & 43.1$^{+7.1}_{-22.4}$ & 0.97$^{+0.52}_{-1.41}$ & 2.23$^{+0.18}_{-0.19}$ & 0.2$^{+1.2}_{-0.2}$ & 1.45 & 32 \\ 
21 & GRB050803 & 6 & 64.8$^{+24.6}_{-10.8}$ & 2.26$^{+0.86}_{-1.37}$ & 9.37$^{+19.37}_{-9.37}$ & 0.3$^{+1.5}_{-0.3}$ & 1.17 & 16 \\ 
22 & GRB050820 & 1 & 4.7$^{+4.7}_{-4.7}$ & 0.17$^{+0.37}_{-0.19}$ & 0.82$^{+0.04}_{-0.04}$ & 3.0$^{+2.1}_{-1.2}$ & 1.12 & 201 \\ 
23 & GRB050822 & 1 & 4.0$^{+7.2}_{-4.0}$ & 0.90$^{+1.14}_{-0.69}$ & 1.91$^{+0.40}_{-0.39}$ & 2.6$^{+2.1}_{-2.5}$ & 0.42 & 25 \\ 
24 & GRB050822 & 2 & 8.9$^{+10.1}_{-2.9}$ & 1.14$^{+0.13}_{-0.82}$ & 2.82$^{+0.33}_{-0.07}$ & 0.4$^{+1.1}_{-0.4}$ & 1.12 & 29 \\ 
25 & GRB050822 & 3 & 1.0$^{+14.4}_{-8.5}$ & 1.06$^{+3.45}_{-0.65}$ & 8.61$^{+1.39}_{-1.39}$ & 0.9$^{+11.9}_{-0.8}$ & 1.93 & 16 \\ 
26 & GRB050904 & 1 & 13.4$^{+3.9}_{-10.2}$ & 0.95$^{+0.94}_{-0.63}$ & 1.75$^{+0.09}_{-0.09}$ & 1.2$^{+14.5}_{-1.2}$ & 0.98 & 180 \\ 
27 & GRB050904 & 4 & 3.3$^{+3.6}_{-2.1}$ & 1.30$^{+0.26}_{-0.35}$ & 9.18$^{+7.00}_{-9.18}$ & 4.2$^{+2.2}_{-1.1}$ & 1.30 & 36 \\ 
28 & GRB050904 & 5 & 1.7$^{+6.1}_{-1.4}$ & 0.29$^{+0.19}_{-0.71}$ & 1.89$^{+0.13}_{-0.14}$ & 0.9$^{+1.3}_{-0.9}$ & 0.89 & 24 \\ 
29 & GRB050904 & 6 & 5.0$^{+5.0}_{-5.0}$ & 1.60$^{+0.78}_{-0.36}$ & 1.81$^{+0.13}_{-0.14}$ & 4.4$^{+11.0}_{-0.9}$ & 1.04 & 21 \\ 
30 & GRB050904 & 7 & 4.0$^{+4.5}_{-1.7}$ & 1.39$^{+0.12}_{-0.56}$ & 2.12$^{+0.38}_{-0.18}$ & 5.8$^{+5.8}_{-3.1}$ & 0.93 & 22 \\ 
31 & GRB050916 & 1 & 97.2$^{+60.1}_{-20.6}$ & 1.00$^{+2.53}_{-0.54}$ & 1.68$^{+0.22}_{-0.33}$ & 1.5$^{+999}_{-1.4}$ & 0.53 & 18 \\ 
32 & GRB050922 & 1 & 28.2$^{+23.2}_{-6.9}$ & 1.80$^{+0.29}_{-1.75}$ & 3.68$^{+0.43}_{-0.84}$ & 0.9$^{+2.4}_{-0.8}$ & 1.03 & 97 \\ 
33 & GRB050922 & 2 & 12.7$^{+40.5}_{-6.3}$ & 1.20$^{+1.14}_{-0.96}$ & 2.54$^{+0.64}_{-7.46}$ & 1.0$^{+999.0}_{-0.9}$ & 0.97 & 43 \\ 
34 & GRB050922 & 3 & 5.1$^{+2.3}_{-2.3}$ & 1.04$^{+0.36}_{-1.01}$ & 2.55$^{+0.19}_{-1.07}$ & 1.6$^{+3.7}_{-0.0}$ & 0.77 & 114 \\ 
35 & GRB051117 & 1 & 10.4$^{+4.0}_{-2.7}$ & 0.87$^{+0.73}_{-0.82}$ & 1.86$^{+0.07}_{-0.07}$ & 1.7$^{+3.6}_{-1.6}$ & 1.10 & 340 \\ 
36 & GRB051117 & 2 & 8.5$^{+2.0}_{-2.0}$ & 0.72$^{+0.19}_{-0.49}$ & 2.18$^{+0.04}_{-0.03}$ & 1.0$^{+0.5}_{-0.9}$ & 0.99 & 316 \\ 
37 & GRB051117 & 3 & 9.4$^{+2.6}_{-3.0}$ & 1.10$^{+0.33}_{-0.49}$ & 2.21$^{+0.06}_{-0.07}$ & 1.0$^{+2.5}_{-0.9}$ & 0.98 & 179 \\ 
38 & GRB051117 & 4 & 11.5$^{+1.7}_{-3.9}$ & 1.10$^{+0.41}_{-0.53}$ & 2.08$^{+0.06}_{-0.06}$ & 1.0$^{+1.4}_{-0.9}$ & 1.04 & 224 \\ 
39 & GRB051117 & 5 & 7.4$^{+3.5}_{-1.2}$ & 1.09$^{+0.34}_{-0.44}$ & 2.44$^{+0.10}_{-0.02}$ & 0.5$^{+0.3}_{-0.4}$ & 1.15 & 182 \\ 
40 & GRB051117 & 6 & 8.8$^{+2.3}_{-3.3}$ & 0.87$^{+0.36}_{-0.50}$ & 2.17$^{+0.05}_{-0.05}$ & 1.1$^{+0.8}_{-1.0}$ & 0.99 & 263 \\ 
41 & GRB051117 & 7 & 7.5$^{+3.2}_{-2.4}$ & 1.09$^{+0.38}_{-0.52}$ & 2.21$^{+0.06}_{-0.07}$ & 1.1$^{+2.7}_{-1.0}$ & 1.10 & 221 \\ 
42 & GRB051227 & 1 & 13.5$^{+12.0}_{-5.4}$ & 0.30$^{+0.20}_{-1.02}$ & 1.83$^{+0.33}_{-1.83}$ & 2.5$^{+6.5}_{-0.1}$ & 0.94 & 22 \\ 
43 & GRB060111 & 1 & 25.8$^{+15.4}_{-6.2}$ & 1.12$^{+0.38}_{-1.79}$ & 2.82$^{+0.14}_{-0.22}$ & 0.9$^{+13.5}_{-0.8}$ & 0.98 & 116 \\ 
44 & GRB060111 & 2 & 27.2$^{+7.5}_{-1.3}$ & 1.73$^{+0.39}_{-0.26}$ & 2.78$^{+0.17}_{-0.15}$ & 0.9$^{+5.5}_{-0.8}$ & 0.94 & 74 \\ 
45 & GRB060111 & 3 & 22.5$^{+5.6}_{-1.4}$ & 1.13$^{+0.09}_{-0.19}$ & 2.22$^{+0.04}_{-0.05}$ & 1.0$^{+0.0}_{-0.9}$ & 0.96 & 295 \\ 
46 & GRB060124 & 1 & 10.5$^{+2.2}_{-3.1}$ & 0.22$^{+0.61}_{-0.62}$ & 1.20$^{+0.02}_{-0.02}$ & 2.0$^{+3.8}_{-0.8}$ & 0.97 & 679 \\ 
47 & GRB060124 & 2 & 8.2$^{+1.2}_{-1.0}$ & 0.14$^{+0.07}_{-0.09}$ & 1.61$^{+0.01}_{-0.01}$ & 0.9$^{+0.2}_{-0.9}$ & 1.06 & 534 \\ 
\enddata
\end{deluxetable}

\begin{deluxetable}{cccccccc}
\tabletypesize{\tiny}
\tablecaption{Properties of exponentially cutoff power law spectral fits to {\it{Gold}} flares \label{tbl:cutoff_table}}
\tablewidth{0pt}
\tablehead{\colhead{Index Number} & \colhead{GRB} & \colhead{Flare} & \colhead{$N_H (10^{20} cm^{-2})$} & \colhead{Photon Index} & \colhead{$E_{cut}$ (keV)} & \colhead{$\chi^2_{red}$} & \colhead{DOF}}
\startdata
1 & GRB050219 & 1 & 40.2$^{+10.6}_{-10.2}$ & 2.68$^{+0.37}_{-0.68}$ & 999 & 1.12 & 37 \\ 
2 & GRB050502 & 1 & 8.7$^{+1.3}_{-1.2}$ & 1.97$^{+0.14}_{-0.14}$ & 7.2$^{+4.4}_{-2.0}$ & 1.36 & 327 \\ 
3 & GRB050502 & 3 & 10.8$^{+5.3}_{-5.5}$ & 2.11$^{+0.24}_{-0.29}$ & 999 & 0.86 & 30 \\ 
4 & GRB050607 & 2 & 22.5$^{+5.2}_{-8.4}$ & 2.40$^{+0.11}_{-0.83}$ & 999 & 0.87 & 32 \\ 
5 & GRB050712 & 1 & 23.9$^{+5.3}_{-3.8}$ & 2.10$^{+0.18}_{-0.16}$ & 999 & 1.60 & 56 \\ 
6 & GRB050712 & 2 & 20.2$^{+9.0}_{-7.8}$ & 3.06$^{+0.55}_{-0.95}$ & 999 & 1.09 & 17 \\ 
7 & GRB050713 & 1 & 46.2$^{+9.4}_{-8.6}$ & 1.82$^{+0.43}_{-0.49}$ & 8.8$^{+999}_{-6.1}$ & 1.02 & 187 \\ 
8 & GRB050713 & 2 & 50.4$^{+13.9}_{-10.9}$ & 3.29$^{+0.65}_{-0.98}$ & 999 & 1.21 & 47 \\ 
9 & GRB050716 & 1 & 23.6$^{+66.7}_{-23.6}$ & 1.18$^{+0.95}_{-1.86}$ & 128.0$^{+128.0}_{-128.0}$ & 0.28 & 55 \\ 
10 & GRB050716 & 2 & 2.3$^{+16.4}_{-2.3}$ & 0.02$^{+2.36}_{-0.66}$ & 0.6$^{+1.5}_{-0.2}$ & 0.32 & 54 \\ 
11 & GRB050724 & 1 & 44.7$^{+3.1}_{-3.0}$ & 1.37$^{+0.14}_{-0.14}$ & 8.7$^{+4.6}_{-2.2}$ & 1.03 & 329 \\ 
12 & GRB050724 & 2 & 48.1$^{+9.3}_{-8.9}$ & 2.32$^{+0.67}_{-0.70}$ & 4.3$^{+4.3}_{-2.3}$ & 0.95 & 53 \\ 
13 & GRB050724 & 3 & 18.0$^{+13.1}_{-10.5}$ & 0.94$^{+0.74}_{-0.71}$ & 4.8$^{+62.5}_{-2.6}$ & 1.25 & 21 \\ 
14 & GRB050726 & 2 & 1.6$^{+19.7}_{-1.6}$ & 0.63$^{+2.25}_{-0.80}$ & 1.1$^{+498.9}_{-0.4}$ & 0.90 & 36 \\ 
15 & GRB050730 & 1 & 10.7$^{+4.6}_{-6.3}$ & 1.61$^{+0.21}_{-0.55}$ & 32.2$^{+32.2}_{-27.6}$ & 1.08 & 57 \\ 
16 & GRB050730 & 2 & 8.4$^{+1.0}_{-0.5}$ & 1.66$^{+0.05}_{-0.05}$ & 999 & 0.94 & 186 \\ 
17 & GRB050730 & 3 & 5.8$^{+0.9}_{-1.1}$ & 1.91$^{+0.08}_{-0.04}$ & 999 & 0.94 & 105 \\ 
18 & GRB050730 & 4 & 7.7$^{+6.0}_{-5.0}$ & 1.58$^{+0.62}_{-0.62}$ & 4.1$^{+235.8}_{-2.1}$ & 0.98 & 80 \\ 
19 & GRB050802 & 1 & 1.8$^{+1.8}_{-1.8}$ & 1.17$^{+0.53}_{-0.64}$ & 2.4$^{+7.0}_{-1.2}$ & 0.93 & 30 \\ 
20 & GRB050803 & 5 & 46.9$^{+10.0}_{-6.8}$ & 2.23$^{+0.24}_{-0.18}$ & 999 & 1.45 & 33 \\ 
21 & GRB050803 & 6 & 83.5$^{+32.5}_{-41.3}$ & 4.14$^{+1.50}_{-3.99}$ & 5.9$^{+5.9}_{-5.4}$ & 1.08 & 17 \\ 
22 & GRB050820 & 1 & 8.0$^{+3.4}_{-2.4}$ & 0.72$^{+0.12}_{-0.14}$ & 33.6$^{+33.6}_{-19.2}$ & 1.13 & 201 \\ 
23 & GRB050822 & 1 & 4.0$^{+13.0}_{-4.0}$ & 0.99$^{+1.02}_{-0.72}$ & 3.3$^{+999}_{-1.7}$ & 0.42 & 26 \\ 
24 & GRB050822 & 2 & 19.3$^{+4.8}_{-8.7}$ & 2.86$^{+0.26}_{-1.05}$ & 275.2$^{+999}_{-273.1}$ & 1.07 & 30 \\ 
25 & GRB050822 & 3 & 13.2$^{+14.8}_{-6.0}$ & 1.42$^{+1.54}_{-0.35}$ & 0.5$^{+0.1}_{-0.3}$ & 0.75 & 17 \\ 
26 & GRB050904 & 1 & 14.4$^{+2.8}_{-2.3}$ & 1.77$^{+0.09}_{-0.15}$ & 496.7$^{+999}_{-488.3}$ & 0.99 & 181 \\ 
27 & GRB050904 & 4 & 3.9$^{+3.6}_{-3.0}$ & 1.37$^{+0.37}_{-0.35}$ & 4.7$^{+7.3}_{-1.8}$ & 1.27 & 37 \\ 
28 & GRB050904 & 5 & 11.8$^{+3.5}_{-6.0}$ & 1.93$^{+0.08}_{-0.56}$ & 120.8$^{+999}_{-999}$ & 0.96 & 25 \\ 
29 & GRB050904 & 6 & 4.2$^{+2.4}_{-2.5}$ & 1.81$^{+0.12}_{-0.27}$ & 495.6$^{+999}_{-485.9}$ & 1.05 & 21 \\ 
30 & GRB050904 & 7 & 5.5$^{+4.0}_{-3.7}$ & 1.59$^{+0.34}_{-0.43}$ & 10.6$^{+10.6}_{-6.7}$ & 0.89 & 23 \\ 
31 & GRB050916 & 1 & 94.8$^{+36.9}_{-40.9}$ & 1.58$^{+0.41}_{-1.32}$ & 37.7$^{+999}_{-35.0}$ & 0.50 & 19 \\ 
32 & GRB050922 & 1 & 47.7$^{+16.4}_{-24.8}$ & 3.92$^{+0.75}_{-3.04}$ & 204.3$^{+999}_{-999}$ & 1.02 & 98 \\ 
33 & GRB050922 & 2 & 19.5$^{+18.7}_{-19.5}$ & 2.57$^{+0.92}_{-3.05}$ & 25.2$^{+999}_{-999}$ & 0.94 & 44 \\ 
34 & GRB050922 & 3 & 9.1$^{+4.1}_{-3.6}$ & 1.72$^{+0.43}_{-0.41}$ & 4.1$^{+7.7}_{-1.6}$ & 0.77 & 115 \\ 
35 & GRB051117 & 1 & 16.2$^{+1.6}_{-1.9}$ & 1.83$^{+0.08}_{-0.16}$ & 62.7$^{+999}_{-47.9}$ & 1.11 & 341 \\ 
36 & GRB051117 & 2 & 16.4$^{+1.3}_{-1.7}$ & 2.19$^{+0.07}_{-0.15}$ & 70.7$^{+999}_{-56.2}$ & 1.04 & 317 \\ 
37 & GRB051117 & 3 & 13.4$^{+1.6}_{-1.8}$ & 2.25$^{+0.07}_{-0.18}$ & 999 & 0.99 & 180 \\ 
38 & GRB051117 & 4 & 14.3$^{+1.4}_{-0.7}$ & 2.12$^{+0.07}_{-0.12}$ & 999 & 1.05 & 225 \\ 
39 & GRB051117 & 5 & 14.0$^{+1.7}_{-1.3}$ & 2.50$^{+0.08}_{-0.10}$ & 999 & 1.25 & 183 \\ 
40 & GRB051117 & 6 & 13.6$^{+2.0}_{-1.9}$ & 1.97$^{+0.19}_{-0.19}$ & 11.2$^{+33.1}_{-4.9}$ & 1.01 & 264 \\ 
41 & GRB051117 & 7 & 12.4$^{+1.2}_{-2.0}$ & 2.25$^{+0.06}_{-0.21}$ & 497.6$^{+999}_{-485.7}$ & 1.11 & 222 \\ 
42 & GRB051227 & 1 & 17.7$^{+11.9}_{-10.2}$ & 0.73$^{+0.63}_{-0.62}$ & 4.2$^{+13.4}_{-1.9}$ & 0.90 & 23 \\ 
43 & GRB060111 & 1 & 38.0$^{+4.3}_{-7.4}$ & 2.84$^{+0.18}_{-0.62}$ & 59.9$^{+999}_{-55.9}$ & 0.98 & 117 \\ 
44 & GRB060111 & 2 & 31.2$^{+4.0}_{-4.0}$ & 2.86$^{+0.17}_{-0.30}$ & 499.9$^{+999}_{-490.9}$ & 0.96 & 75 \\ 
45 & GRB060111 & 3 & 26.5$^{+1.3}_{-0.7}$ & 2.27$^{+0.04}_{-0.06}$ & 500.0$^{+999}_{-441.5}$ & 1.01 & 296 \\ 
46 & GRB060124 & 1 & 17.4$^{+0.8}_{-0.8}$ & 1.14$^{+0.05}_{-0.05}$ & 48.5$^{+106.7}_{-19.1}$ & 0.98 & 680 \\ 
47 & GRB060124 & 2 & 16.4$^{+0.5}_{-0.4}$ & 1.66$^{+0.02}_{-0.02}$ & 499.9$^{+999}_{-364.8}$ & 1.12 & 535 \\ 
\enddata
\end{deluxetable}

\begin{deluxetable}{cccccccc}
\tabletypesize{\tiny}
\tablecaption{Properties of blackbody plus power law spectral fits to {\it{Gold}} flares \label{tbl:bb_table}}
\tablewidth{0pt}
\tablehead{\colhead{Index Number} & \colhead{GRB} & \colhead{Flare} & \colhead{$N_H (10^{20} cm^{-2})$} & \colhead{Photon Index} & \colhead{kT (keV)} & \colhead{$\chi^2_{red}$} & \colhead{DOF}}
\startdata
1 & GRB050219 & 1 & 56.8$^{+55.9}_{-29.1}$ & 2.34$^{+0.75}_{-0.81}$ & 0.1$^{+0.1}_{-0.1}$ & 0.96 & 36 \\ 
2 & GRB050502 & 1 & 6.4$^{+1.3}_{-1.5}$ & 2.06$^{+0.08}_{-0.10}$ & 0.3$^{+0.0}_{-0.0}$ & 1.29 & 326 \\ 
3 & GRB050502 & 3 & 9.5$^{+31.6}_{-9.4}$ & 1.86$^{+2.12}_{-0.36}$ & 0.2$^{+0.6}_{-0.2}$ & 0.84 & 29 \\ 
4 & GRB050607 & 2 & 26.0$^{+25.1}_{-7.8}$ & 2.70$^{+1.98}_{-0.88}$ & 1.1$^{+0.9}_{-1.0}$ & 0.89 & 31 \\ 
5 & GRB050712 & 1 & 31.2$^{+27.6}_{-2.5}$ & 2.52$^{+1.84}_{-0.10}$ & 32.4$^{+167.6}_{-31.3}$ & 1.52 & 55 \\ 
6 & GRB050712 & 2 & 19.5$^{+10.3}_{-5.7}$ & 3.04$^{+0.63}_{-0.39}$ & 0.0$^{+200.0}_{-0.0}$ & 1.16 & 16 \\ 
7 & GRB050713 & 1 & 41.6$^{+11.4}_{-11.7}$ & 1.93$^{+0.34}_{-0.42}$ & 0.5$^{+0.5}_{-0.1}$ & 1.01 & 186 \\ 
8 & GRB050713 & 2 & 117.6$^{+63.7}_{-54.1}$ & 3.71$^{+1.49}_{-1.03}$ & 0.1$^{+0.0}_{-0.0}$ & 1.03 & 46 \\ 
9 & GRB050716 & 1 & 31.3$^{+255.4}_{-31.3}$ & 1.56$^{+8.36}_{-1.56}$ & 7.2$^{+7.2}_{-7.2}$ & 0.29 & 54 \\ 
10 & GRB050716 & 2 & 11.2$^{+11.2}_{-11.2}$ & 3.29$^{+2.98}_{-0.78}$ & 0.3$^{+0.1}_{-0.1}$ & 0.34 & 54 \\ 
11 & GRB050724 & 1 & 41.7$^{+3.8}_{-3.9}$ & 1.58$^{+0.09}_{-0.11}$ & 0.7$^{+0.1}_{-0.1}$ & 1.01 & 328 \\ 
12 & GRB050724 & 2 & 46.2$^{+11.5}_{-13.4}$ & 2.69$^{+0.47}_{-0.57}$ & 0.4$^{+0.3}_{-0.1}$ & 0.96 & 52 \\ 
13 & GRB050724 & 3 & 14.8$^{+14.9}_{-14.8}$ & 1.31$^{+0.25}_{-1.32}$ & 0.7$^{+0.5}_{-0.2}$ & 1.29 & 20 \\ 
14 & GRB050726 & 2 & 2.3$^{+17.5}_{-2.3}$ & 2.03$^{+3.13}_{-1.19}$ & 0.4$^{+0.3}_{-0.1}$ & 0.90 & 35 \\ 
15 & GRB050730 & 1 & 3.2$^{+3.2}_{-3.2}$ & 1.29$^{+0.12}_{-0.15}$ & 0.4$^{+0.1}_{-0.1}$ & 1.00 & 57 \\ 
16 & GRB050730 & 2 & 12.6$^{+2.7}_{-2.3}$ & 2.00$^{+0.21}_{-0.18}$ & 2.8$^{+22.1}_{-0.8}$ & 0.90 & 185 \\ 
17 & GRB050730 & 3 & 9.6$^{+3.0}_{-2.5}$ & 2.28$^{+0.27}_{-0.22}$ & 2.1$^{+3.1}_{-0.6}$ & 0.91 & 104 \\ 
18 & GRB050730 & 4 & 3.1$^{+4.9}_{-3.1}$ & 1.75$^{+0.32}_{-0.35}$ & 0.4$^{+0.1}_{-0.1}$ & 0.95 & 79 \\ 
19 & GRB050802 & 1 & 6.2$^{+29.3}_{-6.2}$ & 2.45$^{+1.69}_{-0.78}$ & 0.7$^{+0.5}_{-0.2}$ & 1.00 & 28 \\ 
20 & GRB050803 & 5 & 40.0$^{+32.8}_{-12.9}$ & 0.71$^{+0.46}_{-0.69}$ & 0.3$^{+0.1}_{-0.1}$ & 1.16 & 32 \\ 
21 & GRB050803 & 6 & 198.7$^{+90.1}_{-87.5}$ & 5.44$^{+2.29}_{-1.41}$ & 0.1$^{+0.0}_{-0.0}$ & 0.98 & 16 \\ 
22 & GRB050820 & 1 & 11.1$^{+2.1}_{-1.8}$ & 0.84$^{+0.04}_{-0.04}$ & 0.0$^{+0.0}_{-0.0}$ & 1.12 & 200 \\ 
23 & GRB050822 & 1 & 2.3$^{+2.3}_{-2.3}$ & 1.32$^{+0.30}_{-0.41}$ & 0.4$^{+0.2}_{-0.1}$ & 0.39 & 26 \\ 
24 & GRB050822 & 2 & 19.2$^{+11.9}_{-14.1}$ & 2.75$^{+0.42}_{-0.08}$ & 0.1$^{+0.1}_{-0.1}$ & 1.09 & 29 \\ 
25 & GRB050822 & 3 & 34.3$^{+29.0}_{-3.0}$ & 4.57$^{+1.71}_{-0.51}$ & 0.0$^{+0.0}_{-0.0}$ & 0.94 & 16 \\ 
26 & GRB050904 & 1 & 14.8$^{+2.5}_{-2.5}$ & 1.79$^{+0.09}_{-0.07}$ & 0.0$^{+0.5}_{-0.0}$ & 0.99 & 180 \\ 
27 & GRB050904 & 4 & 10.4$^{+5.0}_{-3.9}$ & 2.29$^{+0.43}_{-0.27}$ & 1.0$^{+0.2}_{-0.2}$ & 1.29 & 36 \\ 
28 & GRB050904 & 5 & 5.2$^{+6.5}_{-4.2}$ & 1.50$^{+0.34}_{-0.46}$ & 0.3$^{+0.1}_{-0.1}$ & 0.90 & 24 \\ 
29 & GRB050904 & 6 & 6.0$^{+2.6}_{-2.9}$ & 1.97$^{+0.27}_{-0.22}$ & 31.0$^{+31.0}_{-31.0}$ & 1.06 & 20 \\ 
30 & GRB050904 & 7 & 7.6$^{+2.3}_{-3.5}$ & 1.96$^{+0.20}_{-0.24}$ & 0.9$^{+0.9}_{-0.9}$ & 0.91 & 22 \\ 
31 & GRB050916 & 1 & 84.1$^{+56.8}_{-52.3}$ & 1.38$^{+1.37}_{-2.69}$ & 0.5$^{+199.4}_{-0.5}$ & 0.49 & 18 \\ 
32 & GRB050922 & 1 & 17.3$^{+74.6}_{-12.9}$ & 1.86$^{+1.93}_{-2.11}$ & 0.2$^{+0.1}_{-0.2}$ & 0.99 & 97 \\ 
33 & GRB050922 & 2 & 20.0$^{+22.5}_{-11.3}$ & 2.65$^{+1.27}_{-0.27}$ & 199.3$^{+0.7}_{-199.3}$ & 0.97 & 43 \\ 
34 & GRB050922 & 3 & 11.5$^{+4.0}_{-4.3}$ & 2.30$^{+0.24}_{-0.23}$ & 0.5$^{+0.2}_{-0.1}$ & 0.77 & 114 \\ 
35 & GRB051117 & 1 & 14.3$^{+2.6}_{-2.5}$ & 1.77$^{+0.11}_{-0.12}$ & 0.4$^{+0.3}_{-0.1}$ & 1.11 & 340 \\ 
36 & GRB051117 & 2 & 13.2$^{+1.9}_{-2.4}$ & 2.06$^{+0.09}_{-0.13}$ & 0.3$^{+0.0}_{-0.0}$ & 1.01 & 316 \\ 
37 & GRB051117 & 3 & 11.6$^{+3.8}_{-3.0}$ & 2.14$^{+0.16}_{-0.18}$ & 0.3$^{+0.2}_{-0.2}$ & 0.99 & 179 \\ 
38 & GRB051117 & 4 & 11.8$^{+4.7}_{-3.1}$ & 1.99$^{+0.33}_{-0.12}$ & 0.3$^{+199.7}_{-0.3}$ & 1.04 & 224 \\ 
39 & GRB051117 & 5 & 8.4$^{+2.9}_{-2.7}$ & 2.04$^{+0.17}_{-0.18}$ & 0.2$^{+0.0}_{-0.0}$ & 1.08 & 182 \\ 
40 & GRB051117 & 6 & 15.1$^{+2.0}_{-2.7}$ & 2.20$^{+0.12}_{-0.13}$ & 0.5$^{+0.3}_{-0.5}$ & 1.03 & 263 \\ 
41 & GRB051117 & 7 & 10.6$^{+2.5}_{-2.9}$ & 2.16$^{+0.12}_{-0.17}$ & 0.3$^{+0.1}_{-0.3}$ & 1.11 & 221 \\ 
42 & GRB051227 & 1 & 16.3$^{+19.4}_{-12.8}$ & 1.24$^{+0.94}_{-0.67}$ & 0.7$^{+0.7}_{-0.2}$ & 0.95 & 22 \\ 
43 & GRB060111 & 1 & 35.7$^{+26.1}_{-11.5}$ & 2.77$^{+0.56}_{-0.50}$ & 0.3$^{+1.0}_{-0.3}$ & 0.99 & 116 \\ 
44 & GRB060111 & 2 & 45.0$^{+18.2}_{-13.9}$ & 2.86$^{+0.30}_{-0.28}$ & 0.1$^{+0.0}_{-0.0}$ & 0.87 & 74 \\ 
45 & GRB060111 & 3 & 30.5$^{+5.7}_{-4.2}$ & 2.18$^{+0.09}_{-0.09}$ & 0.1$^{+0.0}_{-0.0}$ & 0.93 & 295 \\ 
46 & GRB060124 & 1 & 19.4$^{+0.4}_{-0.5}$ & 1.22$^{+0.01}_{-0.01}$ & 0.0$^{+0.0}_{-0.0}$ & 0.95 & 679 \\ 
47 & GRB060124 & 2 & 13.6$^{+1.2}_{-1.2}$ & 1.53$^{+0.05}_{-0.05}$ & 0.4$^{+0.0}_{-0.0}$ & 1.10 & 534 \\ 
\enddata
\end{deluxetable}

\begin{figure}
\includegraphics[scale=0.4]{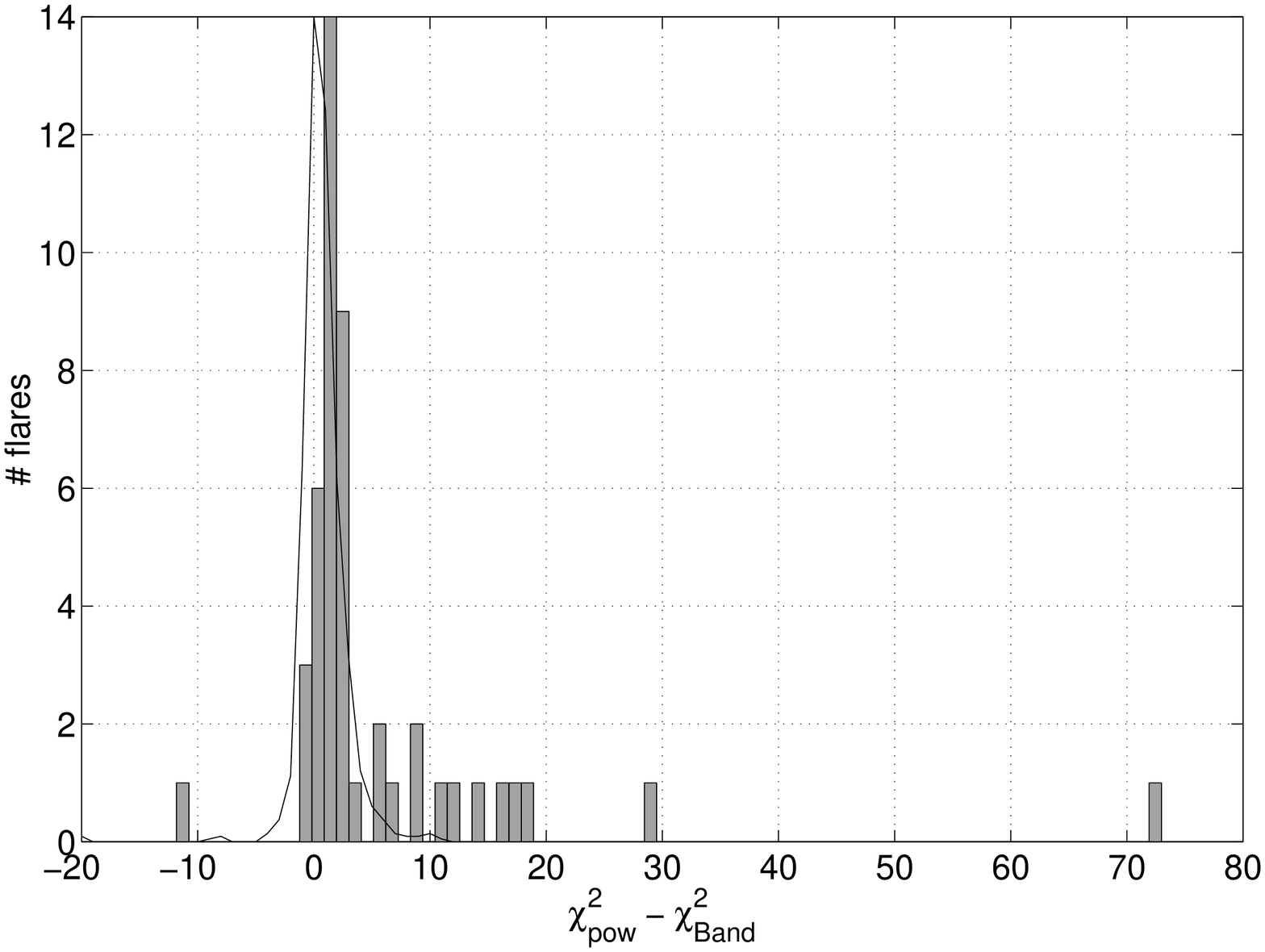}
\caption{Histogram of $\Delta\chi^2$ between the power law fits and Band function fits for all {\it{Gold}} flares.  The histogram represents the real data, while the overlayed line represents the distribution of simulated power law spectra subjected to the same fitting procedure.}
\label{fig:delta_chi2_pow_band_gold}
\end{figure}

It is clear that there are many cases for which a power law provides a satisfactory fit.  However, it is also clear that there are many cases for which a more complex model, such as a Band function, provides a superior fit.  In order to explore the distribution of the change in the quality of fit, a histogram of the $\Delta\chi^2$ between the power law fits and the Band function fits is shown in Figure \ref{fig:delta_chi2_pow_band_gold}.  This histogram shows the $\chi_{pow}^2 - \chi_{Band}^2$ for the 47 {\it{Gold}} flares.  The mean degrees of freedom were 130 and 128 respectively.  For comparison, we have also simulated this same $\Delta\chi^2$ for a fake distribution of power law spectra.  To do this, we simulated 1000 fake spectra that were power laws with spectral photon indices 1.9$\pm$0.3 and the same mean degrees of freedom as the spectra in the flare sample.  These spectra were then fit in the same way that we fit the spectra of the flare sample, and the $\Delta\chi^2$ was calculated.  The resulting $\Delta\chi^2$ distribution from this simulation was plotted as a curve overlaying the histogram of the real data in Figure \ref{fig:delta_chi2_pow_band_gold}.  

As expected, there are many flares that can be fit equally well by both models, but some flares have a better fit using a complex model such as the Band function.  The distribution is skewed to the positive values of $\Delta\chi^2$.  One way to quantify this is to compare the number of GRB flares with a large $\Delta\chi^2$ for both the real data and the simulated power law data.  For the simulated power law data, there are 5 events from 1000 with a $\Delta\chi^2$$>9.0$, so one would expect 0.23 flares in the sample of 47 {\it{Gold}} flares to have a $\Delta\chi^2$$>9.0$ by chance, if the real data had been drawn from a power law distribution.  For the real data, there are 9 GRB flares from the sample of 47 {\it{Gold}} flares that have a $\Delta\chi^2$$>9.0$.  Therefore, it is unlikely that this sample was drawn from a simple power law spectral distribution, and it is clear that Band functions sometimes provide a superior fit.  However, it is worth mentioning that a power law can provide a reasonable fit to many flares.

\subsection{Fluence of Flares}

The fluence of a flare is defined as the flux of the flare, found using the spectral fits described above, integrated over the duration of the flare from $t_{start}$ to $t_{stop}$ in the 0.2--10 keV energy band.  Table \ref{tbl:fluence_table} shows the fluences for both the power law and the Band function fits to the flares.  Since the spectral fits provided no compelling evidence for using the thermal model or the exponential cutoff model, for the remainder of this paper, we restrict ourselves to the standard spectral models used for GRB afterglows and prompt emission, namely the simple power law and the Band function.  The quoted error bars are 1$\sigma$, and they include the error due to the uncertainty in the underlying lightcurve contribution to the fluence.  In some cases this latter source of error is large and dominates the error from the spectral fit itself.  The calculated fluence values for all flares (not just the {\it{Gold}} flares) have been reported, even in cases for which there are very few degrees of freedom.  As a result, some of the fluence values are poorly constrained, as reflected by the error bars.  

These reported fluence values do not include the contribution from the power law component of the spectral model that was used to approximate the underlying afterglow lightcurve contribution to the flare spectrum.  In other words, the fluence values in Column 6 of Table \ref{tbl:fluence_table} include only the contribution from the Band function component of the spectral model fit to the flare data, excluding the frozen power law component from the underlying component.  The fluence values in Column 3 of Table \ref{tbl:fluence_table} include only the contribution from the unfrozen flare power law component of the spectral model fit to the flare data, once again excluding the underlying component.  This is an important point to stress since most previous papers that quote a fluence for flares actually quote the entire fluence under the lightcurve.  This practice is misleading because the underlying afterglow lightcurve sometimes contributes a large, and difficult to constrain, fraction of the total fluence.  The reported fluence values are also corrected for effects due to incomplete light curves for some flares.  For instance, if the tail end of a flare happened to be interrupted by an orbital gap or the South Atlantic Anomaly, the incomplete flare light curve would be extrapolated until it intersected with the extrapolation of the underlying light curve.  In these cases, the error on this extrapolation was factored into the error on the fluence.

The fluence calculations were done in the observed XRT energy band, which is 0.2--10 keV.  Since the X-ray flares emit the bulk of their energy in this band, we consider this to be the most reasonable approach.  This fact is supported by the Band function spectral fits that result in X-ray peak energies, and it is supported by the fact that the X-ray flares are typically weak to undetectable by higher energy and lower energy instruments such as BAT and UVOT.  We computed the effect of extrapolating the typical flare spectrum into the higher energy band typically reported for {\it{Swift}}-BAT bursts.  The median spectral parameters for {\it{Gold}} flares that had a reasonable spectral fit were $\alpha=1.06$, $\beta=2.21$, and $E_0=1.02$ keV; where these refer to the Band function lower energy photon index, higher energy photon index, and e-folding energy, respectively.  From these values, it was found that extending the energy range from 0.2--150 keV added only 1.4\% to the fluence relative to the reported 0.2--10 keV value.  This is insignificant compared to the error bars.

\begin{figure}
\includegraphics[scale=0.7]{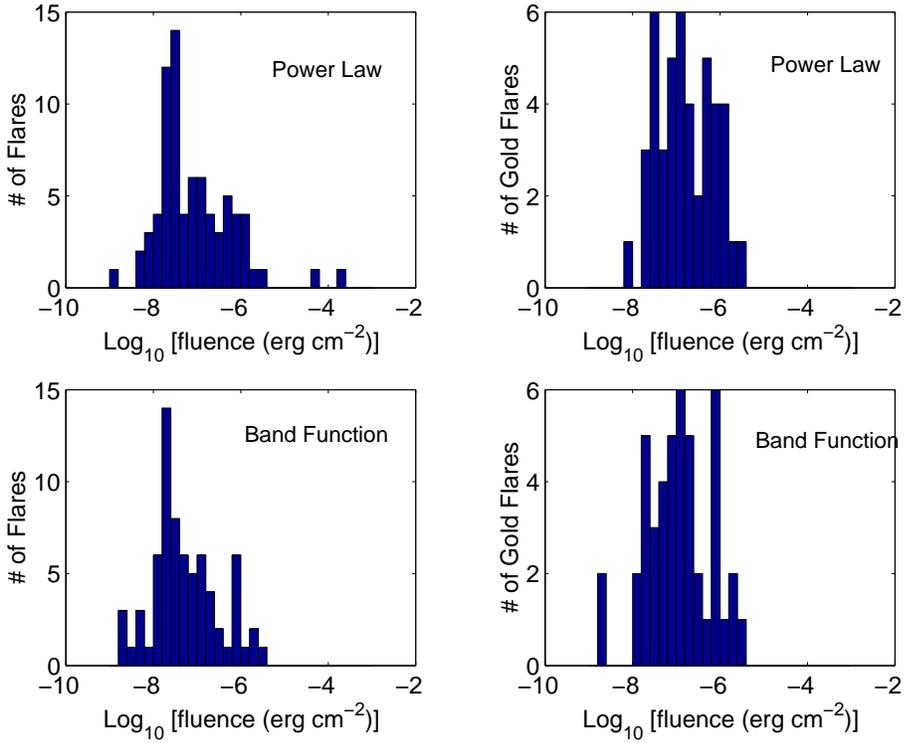}
\caption{Unabsorbed 0.2--10.0 keV fluence distribution of flares.  The two panels on the left are for all flares that had a convergent spectral fit.  The two panels on the right are for {\it{Gold}} flares that have $>15$ DOF in the spectral fit and $\chi^{2}_{red}<1.5$.  Fluence derived from both power law fits (top) and Band function fits (bottom) are shown.}
\label{fig:fluence_distribution}
\end{figure}

The overall distribution of flare fluences is shown in Figure \ref{fig:fluence_distribution} for the absorbed power law model and the absorbed Band function model.  Once again, these fluences are for just the flare component.  The two distributions plotted on the right are for the 47 {\it{Gold}} flares that have $>15$ DOF in the spectral fit and $\chi^{2}_{red}<1.5$, while the two distributions plotted on the left include all spectral fits.  Fluence derived from both power law fits (top) and Band function fits (bottom) are shown.  The mean 0.2--10.0 keV fluence (unabsorbed) of our sample of flares, derived using Band function fits, is $2.4\times10^{-7}$ erg cm$^{-2}$.  There is no evidence of a bimodal distribution, which might arise if flares came from multiple processes.

\begin{deluxetable}{cccccccc}
\tabletypesize{\tiny}
\tablecaption{Fluence of Flares \label{tbl:fluence_table}}
\tablewidth{0pt}
\tablehead{\colhead{GRB} & \colhead{Flare} & \colhead{$Fluence (10^{-7} erg~cm^{-2})$} & \colhead{$\chi^2_{red}$} & \colhead{DOF} & \colhead{$Fluence (10^{-7} erg~cm^{-2})$} & \colhead{$\chi^2_{red}$} & \colhead{DOF} \\ \colhead{} & \colhead{} & \colhead{Power Law} & \colhead{} & \colhead{} & \colhead{Band Function} & \colhead{} & \colhead{}}
\startdata
GRB050219 & 1 & 0.70$^{+999.00}_{-999.00}$ & 1.09 & 38 & 0.38$^{+0.37}_{-0.38}$ & 1.12 & 36 \\ 
GRB050406 & 1 & 0.21$^{+0.02}_{-0.07}$ & 1.78 & 7 & 0.18$^{+0.16}_{-2.02}$ & 2.43 & 6 \\ 
GRB050421 & 1 & 0.21$^{+0.17}_{-999.00}$ & 0.53 & 10 & 0.28$^{+0.27}_{-0.27}$ & 0.64 & 8 \\ 
GRB050502 & 1 & 12.99$^{+0.19}_{-0.20}$ & 1.41 & 328 & 8.30$^{+8.30}_{-0.44}$ & 1.20 & 326 \\ 
GRB050502 & 2 & 0.24$^{+0.17}_{-0.07}$ & 1.21 & 8 & 0.16$^{+0.14}_{-0.16}$ & 1.54 & 6 \\ 
GRB050502 & 3 & 0.90$^{+0.11}_{-999.00}$ & 0.83 & 31 & 0.81$^{+0.79}_{-0.31}$ & 0.85 & 29 \\ 
GRB050607 & 1 & 0.20$^{+999.00}_{-999.00}$ & 0.73 & 9 & 0.20$^{+0.14}_{-1.12}$ & 0.83 & 7 \\ 
GRB050607 & 2 & 1.09$^{+999.00}_{-999.00}$ & 0.84 & 33 & 0.76$^{+0.76}_{-0.35}$ & 0.87 & 31 \\ 
GRB050712 & 1 & 1.51$^{+999.00}_{-999.00}$ & 1.57 & 57 & 1.57$^{+3.26}_{-2.93}$ & 1.47 & 55 \\ 
GRB050712 & 2 & 0.40$^{+999.00}_{-999.00}$ & 1.03 & 18 & 0.26$^{+0.99}_{-2.44}$ & 0.98 & 16 \\ 
GRB050712 & 3 & 0.35$^{+999.00}_{-999.00}$ & 0.73 & 10 & 0.18$^{+0.18}_{-0.09}$ & 0.86 & 8 \\ 
GRB050712 & 4 & --- & --- & --- & --- & --- & --- \\ 
GRB050713 & 1 & 3.14$^{+999.00}_{-999.00}$ & 1.02 & 188 & 2.38$^{+1.76}_{-5.87}$ & 1.02 & 186 \\ 
GRB050713 & 2 & 1.55$^{+999.00}_{-999.00}$ & 1.19 & 48 & 0.46$^{+999.00}_{-686.82}$ & 1.24 & 46 \\ 
GRB050714 & 1 & 1790.20$^{+2150.30}_{-2105.80}$ & 1.68 & 15 & 0.42$^{+0.61}_{-32.55}$ & 6.28 & 13 \\ 
GRB050716 & 1 & 0.19$^{+0.17}_{-0.07}$ & 0.28 & 56 & 0.22$^{+999.00}_{-999.00}$ & 0.28 & 54 \\ 
GRB050716 & 2 & 0.07$^{+0.47}_{-0.69}$ & 0.36 & 55 & 0.02$^{+0.06}_{-0.06}$ & 0.34 & 53 \\ 
GRB050724 & 1 & 0.81$^{+0.01}_{-0.01}$ & 1.05 & 330 & 2.11$^{+1.08}_{-1.08}$ & 1.01 & 328 \\ 
GRB050724 & 2 & 0.31$^{+0.27}_{-0.27}$ & 0.95 & 54 & 0.32$^{+0.48}_{-0.48}$ & 0.97 & 52 \\ 
GRB050724 & 3 & 1.29$^{+0.27}_{-3.04}$ & 1.23 & 22 & 1.28$^{+0.15}_{-0.23}$ & 1.30 & 20 \\ 
GRB050726 & 1 & 0.14$^{+999.00}_{-999.00}$ & 0.73 & 12 & 0.05$^{+999.00}_{-999.00}$ & 0.94 & 10 \\ 
GRB050726 & 2 & 0.26$^{+999.00}_{-999.00}$ & 0.93 & 37 & 0.14$^{+999.00}_{-999.00}$ & 0.92 & 35 \\ 
GRB050730 & 1 & 0.47$^{+0.46}_{-0.54}$ & 1.06 & 58 & 0.35$^{+0.17}_{-0.17}$ & 1.03 & 57 \\ 
GRB050730 & 2 & 2.15$^{+0.36}_{-0.36}$ & 0.94 & 187 & 1.78$^{+0.29}_{-0.29}$ & 0.91 & 185 \\ 
GRB050730 & 3 & 1.03$^{+0.30}_{-0.30}$ & 0.93 & 106 & 0.75$^{+0.22}_{-0.22}$ & 0.84 & 104 \\ 
GRB050730 & 4 & 1.73$^{+1.72}_{-1.69}$ & 0.98 & 81 & 1.06$^{+1.25}_{-1.26}$ & 0.96 & 79 \\ 
GRB050802 & 1 & 0.20$^{+0.33}_{-0.31}$ & 0.97 & 30 & 0.02$^{+0.07}_{-0.07}$ & 0.97 & 29 \\ 
GRB050803 & 1 & 0.30$^{+999.00}_{-999.00}$ & 0.79 & 13 & 0.20$^{+0.12}_{-0.10}$ & 0.93 & 11 \\ 
GRB050803 & 2 & 2.97$^{+999.00}_{-999.00}$ & 1.10 & 14 & 0.30$^{+999.00}_{-999.00}$ & 1.28 & 12 \\ 
GRB050803 & 3 & 0.38$^{+0.03}_{-999.00}$ & 1.34 & 13 & 0.28$^{+0.28}_{-0.06}$ & 1.52 & 11 \\ 
GRB050803 & 4 & 418.24$^{+999.00}_{-999.00}$ & 1.20 & 10 & 0.05$^{+999.00}_{-999.00}$ & 1.58 & 8 \\ 
GRB050803 & 5 & 0.20$^{+2.65}_{-2.65}$ & 1.41 & 34 & 0.10$^{+1.21}_{-1.21}$ & 1.45 & 32 \\ 
GRB050803 & 6 & 0.29$^{+0.89}_{-0.88}$ & 1.02 & 18 & 999.00$^{+999.00}_{-999.00}$ & 1.17 & 16 \\ 
GRB050814 & 1 & 0.04$^{+999.00}_{-999.00}$ & 0.29 & 2 & 0.02$^{+999.00}_{-999.00}$ & 999.00 & 999 \\ 
GRB050814 & 2 & 0.05$^{+999.00}_{-999.00}$ & 0.92 & 6 & 0.04$^{+0.02}_{-999.00}$ & 1.32 & 4 \\ 
GRB050819 & 1 & 0.19$^{+2.76}_{-2.76}$ & 0.50 & 6 & 0.18$^{+2.25}_{-2.39}$ & 0.75 & 5 \\ 
GRB050819 & 2 & 0.10$^{+0.09}_{-0.13}$ & 1.66 & 2 & 999.00$^{+999.00}_{-999.00}$ & 999.00 & 999 \\ 
GRB050820 & 1 & 6.89$^{+108.96}_{-108.96}$ & 1.13 & 202 & 6.81$^{+106.91}_{-106.91}$ & 1.12 & 201 \\ 
GRB050822 & 1 & 0.29$^{+0.03}_{-0.04}$ & 0.44 & 27 & 0.24$^{+0.24}_{-0.01}$ & 0.42 & 25 \\ 
GRB050822 & 2 & 0.95$^{+0.08}_{-0.09}$ & 1.04 & 31 & 0.42$^{+8.48}_{-999.00}$ & 1.12 & 29 \\ 
GRB050822 & 3 & 2.22$^{+41.92}_{-41.92}$ & 1.06 & 18 & 0.17$^{+2.10}_{-2.10}$ & 1.93 & 16 \\ 
GRB050904 & 1 & 2.51$^{+999.00}_{-999.00}$ & 0.98 & 182 & 2.38$^{+2.24}_{-0.17}$ & 0.98 & 180 \\ 
GRB050904 & 2 & 0.27$^{+999.00}_{-999.00}$ & 0.86 & 11 & 0.16$^{+0.08}_{-0.05}$ & 1.03 & 9 \\ 
GRB050904 & 3 & 0.11$^{+999.00}_{-999.00}$ & 1.38 & 6 & 0.10$^{+0.08}_{-1.70}$ & 2.06 & 4 \\ 
GRB050904 & 4 & 0.88$^{+14.49}_{-14.57}$ & 1.31 & 38 & 0.85$^{+13.91}_{-13.91}$ & 1.30 & 36 \\ 
GRB050904 & 5 & 0.95$^{+22.64}_{-22.69}$ & 0.93 & 26 & 1.07$^{+27.60}_{-27.60}$ & 0.89 & 24 \\ 
GRB050904 & 6 & 0.60$^{+17.74}_{-17.99}$ & 1.00 & 22 & 0.57$^{+15.73}_{-15.73}$ & 1.04 & 21 \\ 
GRB050904 & 7 & 0.40$^{+7.53}_{-7.78}$ & 0.86 & 24 & 0.41$^{+7.25}_{-7.25}$ & 0.93 & 22 \\ 
GRB050908 & 1 & 0.26$^{+999.00}_{-0.10}$ & 1.22 & 5 & 0.09$^{+0.08}_{-1.95}$ & 1.79 & 4 \\ 
GRB050908 & 2 & 0.23$^{+0.03}_{-0.04}$ & 0.56 & 14 & 0.20$^{+0.17}_{-0.97}$ & 0.85 & 13 \\ 
GRB050915 & 1 & 0.41$^{+999.00}_{-999.00}$ & 0.84 & 16 & 0.27$^{+0.27}_{-999.00}$ & 0.93 & 14 \\ 
GRB050916 & 1 & 1.30$^{+0.70}_{-999.00}$ & 0.47 & 20 & 1.22$^{+0.04}_{-0.04}$ & 0.53 & 18 \\ 
GRB050922 & 1 & 4.80$^{+999.00}_{-999.00}$ & 1.01 & 99 & 0.74$^{+999.00}_{-25.18}$ & 1.03 & 97 \\ 
GRB050922 & 2 & 0.30$^{+999.00}_{-999.00}$ & 0.92 & 45 & 0.17$^{+999.00}_{-999.00}$ & 0.97 & 43 \\ 
GRB050922 & 3 & 4.57$^{+999.00}_{-999.00}$ & 0.82 & 116 & 2.78$^{+2.75}_{-999.00}$ & 0.77 & 114 \\ 
GRB051006 & 1 & 0.35$^{+0.35}_{-0.04}$ & 0.86 & 6 & 0.21$^{+0.20}_{-999.00}$ & 1.59 & 4 \\ 
GRB051006 & 2 & 0.11$^{+1.92}_{-1.92}$ & 2.04 & 3 & 0.11$^{+1.87}_{-1.87}$ & 6.11 & 1 \\ 
GRB051006 & 3 & 0.30$^{+0.14}_{-999.00}$ & 0.75 & 8 & 0.24$^{+0.24}_{-0.24}$ & 0.98 & 6 \\ 
GRB051016 & 1 & 0.18$^{+0.14}_{-999.00}$ & 1.48 & 1 & 999.00$^{+999.00}_{-999.00}$ & 999.00 & 999 \\ 
GRB051117 & 1 & 20.60$^{+23.04}_{-23.04}$ & 1.11 & 342 & 19.08$^{+22.70}_{-19.05}$ & 1.10 & 340 \\ 
GRB051117 & 2 & 14.24$^{+35.16}_{-35.16}$ & 1.04 & 318 & 11.01$^{+25.35}_{-24.04}$ & 0.99 & 316 \\ 
GRB051117 & 3 & 4.83$^{+56.05}_{-56.05}$ & 0.98 & 181 & 4.20$^{+43.98}_{-43.88}$ & 0.98 & 179 \\ 
GRB051117 & 4 & 7.20$^{+72.04}_{-72.04}$ & 1.04 & 226 & 6.60$^{+61.60}_{-61.41}$ & 1.04 & 224 \\ 
GRB051117 & 5 & 4.91$^{+44.67}_{-44.67}$ & 1.24 & 184 & 3.66$^{+28.55}_{-29.01}$ & 1.15 & 182 \\ 
GRB051117 & 6 & 10.15$^{+86.29}_{-86.29}$ & 1.03 & 265 & 8.22$^{+61.79}_{-61.45}$ & 0.99 & 263 \\ 
GRB051117 & 7 & 8.40$^{+244.17}_{-244.17}$ & 1.11 & 223 & 7.31$^{+184.67}_{-184.60}$ & 1.10 & 221 \\ 
GRB051210 & 1 & 1.00$^{+999.00}_{-999.00}$ & 1.75 & 7 & 0.05$^{+0.02}_{-0.02}$ & 3.51 & 4 \\ 
GRB051227 & 1 & 0.28$^{+0.05}_{-0.05}$ & 0.93 & 24 & 0.20$^{+0.02}_{-0.03}$ & 0.94 & 22 \\ 
GRB060108 & 1 & 0.02$^{+999.00}_{-999.00}$ & 0.29 & 2 & 999.00$^{+999.00}_{-999.00}$ & 999.00 & 999 \\ 
GRB060108 & 2 & 0.70$^{+0.50}_{-999.00}$ & 0.60 & 7 & 0.46$^{+0.34}_{-999.00}$ & 0.80 & 5 \\ 
GRB060109 & 1 & 0.19$^{+0.13}_{-999.00}$ & 0.76 & 16 & 0.32$^{+0.30}_{-0.28}$ & 0.66 & 14 \\ 
GRB060111 & 1 & 4.65$^{+999.00}_{-999.00}$ & 0.98 & 118 & 2.15$^{+4.52}_{-999.00}$ & 0.98 & 116 \\ 
GRB060111 & 2 & 2.05$^{+999.00}_{-999.00}$ & 0.95 & 76 & 1.39$^{+4.13}_{-999.00}$ & 0.94 & 74 \\ 
GRB060111 & 3 & 9.15$^{+999.00}_{-999.00}$ & 1.00 & 297 & 7.20$^{+7.20}_{-1.46}$ & 0.96 & 295 \\ 
GRB060115 & 1 & 0.20$^{+999.00}_{-999.00}$ & 1.06 & 15 & 0.20$^{+0.15}_{-999.00}$ & 0.86 & 13 \\ 
GRB060124 & 1 & 27.13$^{+0.39}_{-0.39}$ & 0.98 & 681 & 33.73$^{+0.47}_{-0.48}$ & 0.97 & 679 \\ 
GRB060124 & 2 & 12.40$^{+0.27}_{-0.26}$ & 1.11 & 536 & 16.80$^{+0.35}_{-0.36}$ & 1.06 & 534 \\  
\enddata
\end{deluxetable}

\section{Flare Fluence versus Prompt Fluence}

\begin{figure}
\includegraphics[scale=0.5]{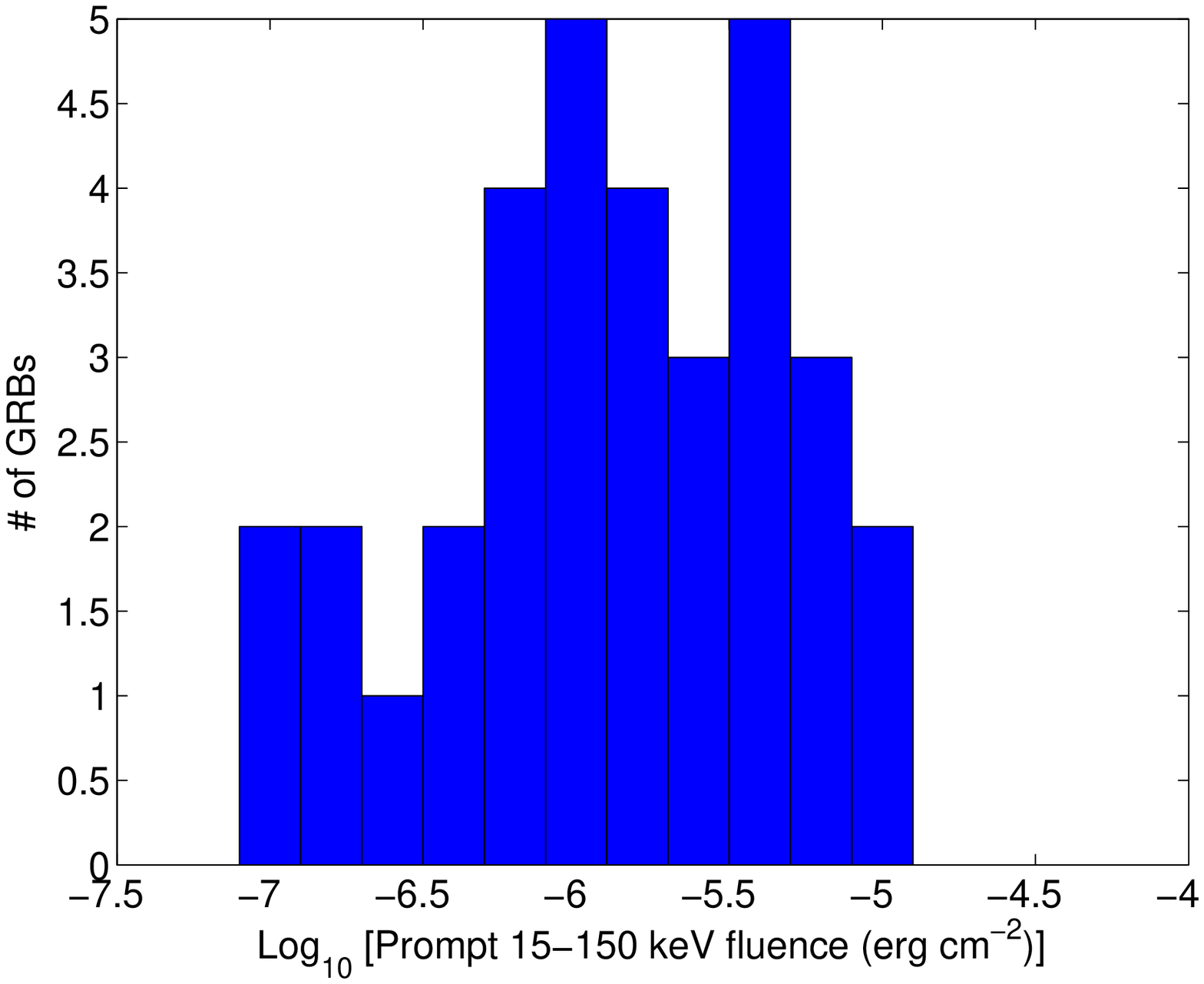}
\caption{Prompt emission 15--150 keV fluence distribution of GRBs that are in this sample of flaring GRBs.}
\label{fig:prompt_fluence_distribution}
\end{figure}

\begin{figure}
\includegraphics[scale=0.5]{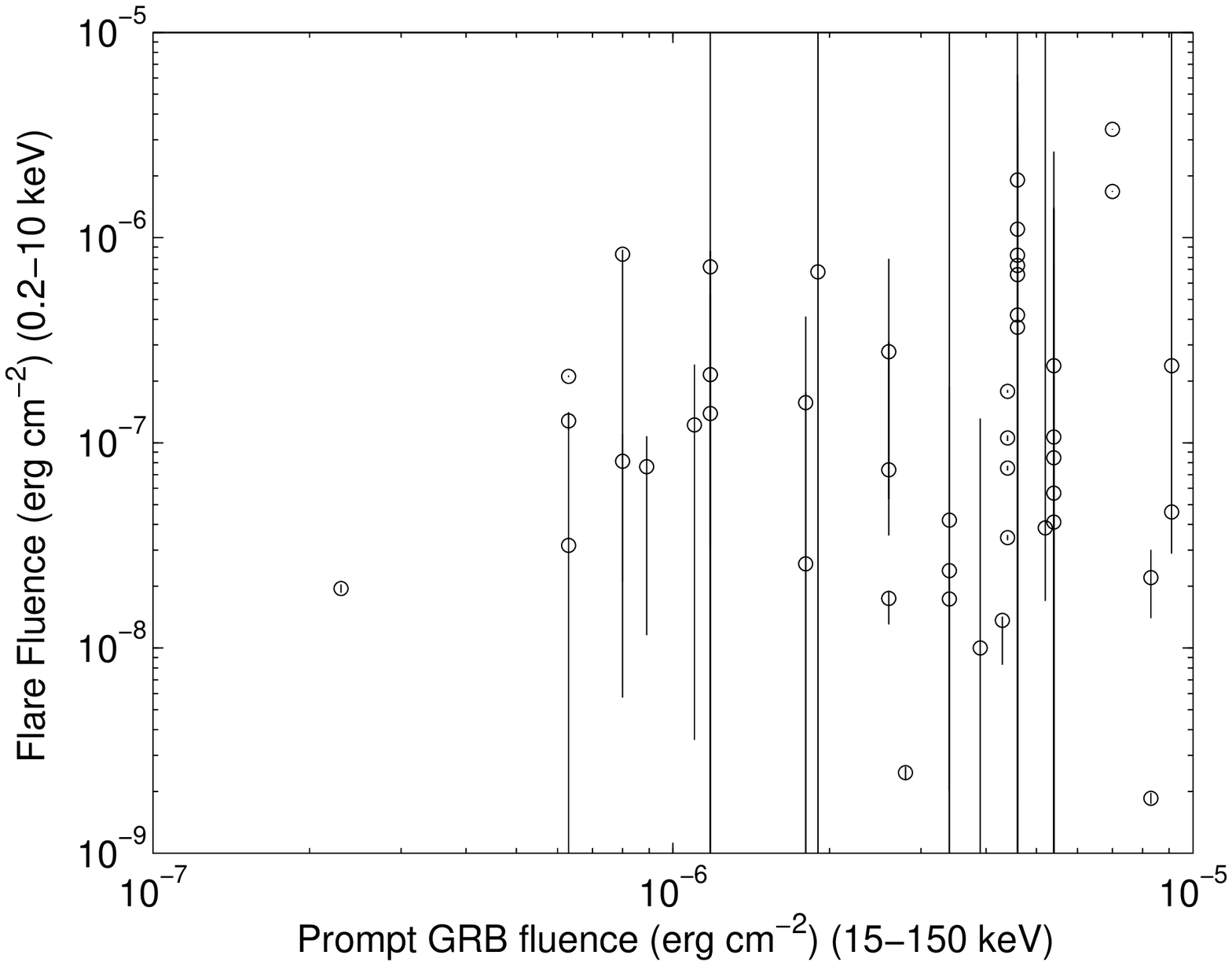}
\caption{Flare fluence in the 0.2--10 keV band (derived using a Band function) plotted as a function of the prompt GRB fluence in the 15--150 keV band.}
\label{fig:flare_vs_prompt_fluence_gold}
\end{figure}

For the purpose of comparison, the distribution of prompt emission fluences for this sample of 33 GRBs is shown in Figure \ref{fig:prompt_fluence_distribution}.  The mean prompt fluence is $2.4\times10^{-6}$ erg cm$^{-2}$, with a standard deviation of $2.5\times10^{-6}$ erg cm$^{-2}$.  The flare fluence in the 0.2--10 keV band, where its energy peaks, is approximately a factor of 10 less than the mean fluence of the prompt GRB emission (calculated in the 15-150 keV band), measured by {\it{Swift}}-BAT.  However, the distributions do overlap.  In at least one case (GRB 050502b), the fluence in a single flare matches the fluence in the prompt emission from the GRB that spawned that flare \citep{fal06}.  The flare fluence in the X-ray band (0.2--10 keV) is plotted as a function of the prompt emission fluence in the 15--150 keV band in Figure \ref{fig:flare_vs_prompt_fluence_gold}.  The instrument bands and the peak energies of the flares and the GRB prompt emission, respectively, have determined the bands over which we have evaluated this fluence.  This is the most reasonable approach in the absence of more refined measurements of all spectral parameters.  However, it should be noted that a correction to bolometric fluence could add significant fluence from lower energies (below 0.2 keV for flares and below 15 keV for GRB prompt emission).  Based on the reasoning in the previous section, it is clear that extending the energy band to higher energies produces only insignificant effects.

\section{Flare Properties versus Underlying Afterglow Properties}

\begin{figure}
\includegraphics[scale=0.5]{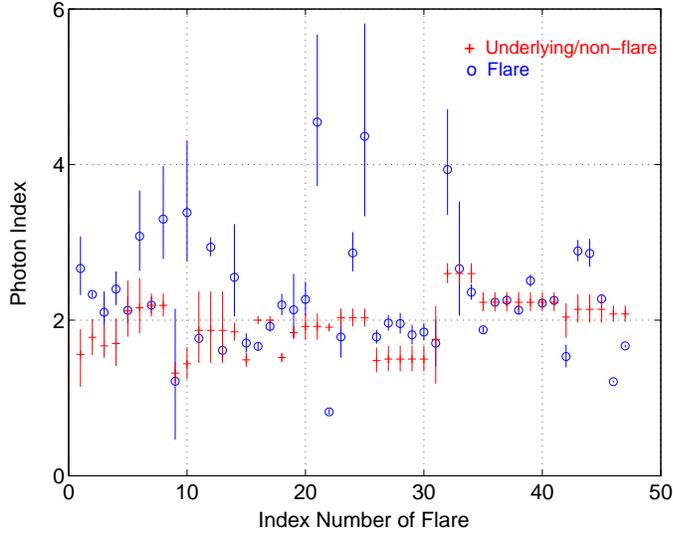}
\caption{Comparison of the power law spectral photon indices for the underlying afterglows and for the flares}
\label{fig:Flare_vs_Underlying_Index}
\end{figure}

Based on rapid temporal properties, repetitive flares, and spectral changes during flares, past studies of individual flaring GRBs have argued that at least the flares in question were due to internal GRB engine properties, as opposed to afterglow related processes.  This idea is further strengthened by the rapid rises and decays seen in the sample of flares presented in Paper I.  In Figure \ref{fig:Flare_vs_Underlying_Index}, we compare the photon index of the power law fit to the underlying afterglow data to the power law photon index of the flare.  The flare power laws have a wider distribution of spectral indices than that of the underlying afterglows.

\section{Temporal Evolution of Flare Properties}

From studies of a few individual GRBs with several strong flares, temporal evolution of spectral properties has been explored.  Spectral evolution in individual bright flares has been seen in GRB 050406 \citep{romano06}, GRB 050502B \citep{fal06}, GRB 050607 \citep{pag06}, GRB 060714 \citep{mor07}, GRB 050822 \citep{god07}, as well as several others.  In all cases, spectral hardening was observed at the onset of the flare, followed by spectral softening as the flare peaked and decayed.  \citet{kri07} discuss temporal evolution of spectral properties from flare to flare in a sequence of flares seen in GRB 060714, which show a decrease in $E_{peak}$ as a function of the time of the flare.  \citet{but07} has observed the same trend in his study of several bright X-ray flares.

\begin{figure}
\includegraphics[scale=0.5]{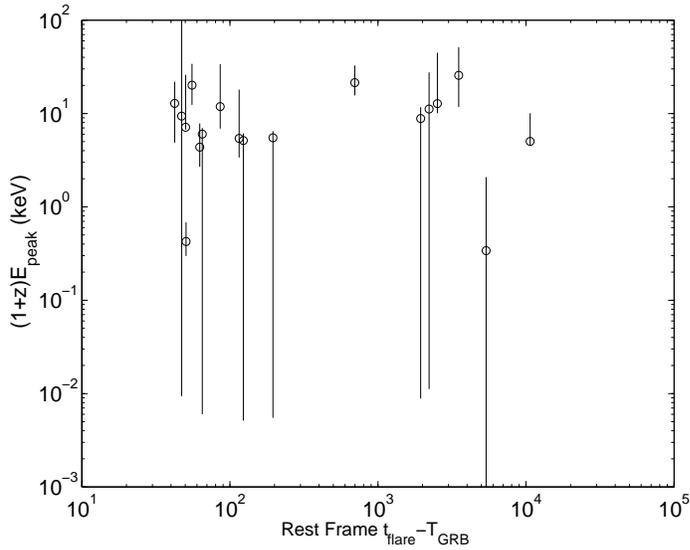}
\caption{Redshift corrected peak energy of {\it{Gold}} flares as a function of rest frame flare time relative to prompt $T_{0}$.  This plot contains all flares irrespective of (and unscaled for) prompt emission $E_{peak}$.}
\label{fig:Epeak_Tflare_gold}
\end{figure}

In Figure \ref{fig:Epeak_Tflare_gold}, we investigate this $E_{peak}$ versus time relationship for this large sample of flares.  We have plotted the values for all of the {\it{Gold}} flares that have a known redshift.  The time axis is the flare time relative to the GRB prompt $T_{0}$ in the burst reference frame.  The results have not been scaled by the prompt GRB $E_{peak}$ or relative to one another in any way.  As a result, this is merely a test of whether or not an absolute relationship exists, independent of the individual GRB or flare parameters.  There is no clear overall relationship present in the data.  Due to the potential for unknown scaling from burst to burst, the spectral softening apparent in individual bursts would not necessarily be apparent when an ensemble of flares from many GRBs is plotted together, as we have done in this case. 

\begin{figure}
\includegraphics[scale=0.5]{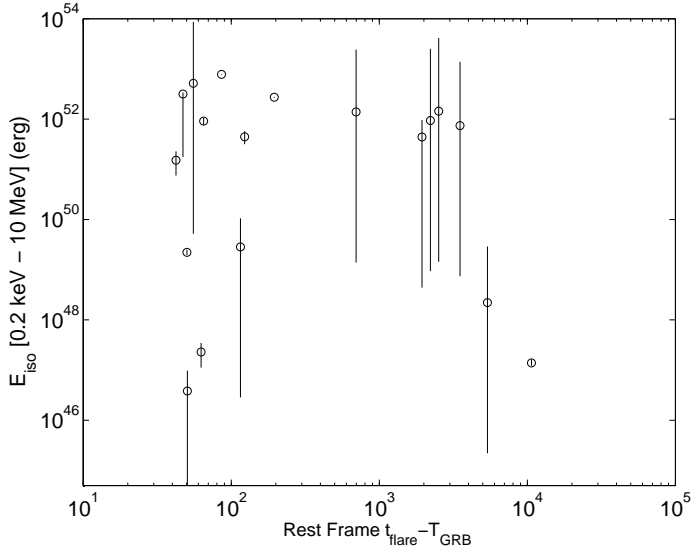}
\caption{$E_{iso}$ for {\it{Gold}} flares as a function of rest frame flare time relative to prompt $T_{0}$.  $E_{iso}$ is k-corrected and is calculated in the 0.2 keV to 10 MeV band.  This plot contains all flares irrespective of (and unscaled for) prompt emission properties.}
\label{fig:Eiso_Tflare_gold}
\end{figure}

In Figure \ref{fig:Eiso_Tflare_gold}, we attempt to investigate the dependence of total energy release on flare time for this sample of flares.  Once again, we have restricted ourselves to plotting the values for all of the {\it{Gold}} flares that have a known redshift.  The time axis is the flare time relative to the the GRB prompt $T_{0}$ in the burst reference frame.  There is no clear relationship in the data when viewing all of the flares as a single sample as we have done here.  Of course, it is still possible that such a temporal relationship exists for the flares in an individual GRB, but a scaling factor (likely to be dependent on prompt GRB parameters) would need to be applied to each GRB and the associated set of flares to see this effect when plotting flare parameters from many GRBs.

\section{E$_{peak}$ versus E$_{iso}$}

Although it is clear that the flares typically peak in the X-ray band, the limited bandwidth of the study presented here (0.3 to 10 keV) makes it difficult to constrain their peak energies well.  However, this is a topic of considerable importance since prompt GRB emission has shown evidence of an empirical relationship between the peak energy of the spectral energy distribution and the total energy in the jet, as well as the observed timescales of the jet emission \citep{ghi05,ama06,lia06,fir06,tho07}.  So, in spite of the narrow/unconstraining bandwidth, we have attempted to explore this relationship by looking at the relationship between the Band function E$_{peak}$ and E$_{iso}$ in this band.  E$_{peak}$ is the peak energy of the Band function spectrum for the flare, which corresponds to $(2+\alpha)E_{0}$, where $E_0$ is the e-folding energy that is obtained from a Band function spectral fit.  In this work, E$_{iso}$ is defined as the isotropic equivalent energy released during the GRB flare in the 0.2 keV to 10 MeV band, assuming a Band function spectrum derived in the 0.2 to 10 keV band, where the observed flare spectra typically peak.  This is calculated as:
\begin{equation}
E_{iso}=k{\times}\frac{4{\pi}d_{lum}^2}{(1+z)}{\times}[S_{obs}]
\end{equation}
where $S_{obs}$ is the unabsorbed fluence seen by the observer in the 0.2--10 keV band, $z$ is the redshift, $k$ is the correction factor from the observed 0.2--10 keV band to the co-moving 0.2 keV to 10 MeV band, and $d_{lum}$ is the luminosity distance calculated using a flat $\Lambda$ dominated universe with $\Omega_{M}=0.31$, $\Omega_{\Lambda}=0.69$, and $H_0=70$ km s$^{-1}$ Mpc$^{-1}$.  The k-correction factor is described in detail in \citet{blo01}.

\begin{figure}
\includegraphics[scale=0.5]{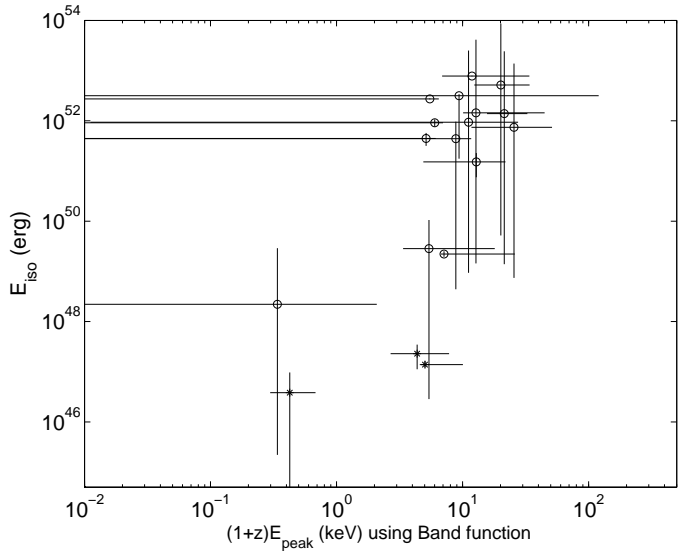}
\caption{Exploration of Band function fit for E$_{peak}$ relationship with E$_{iso}$ for flare emission.  E$_{iso}$ has been k corrected into the co-moving 0.2 keV to 10.0 MeV band.  Only the fluence from the flare itself (i.e. underlying afterglow emission subtracted) was included in the calculation of E$_{iso}$.  The three data points with the lowest E$_{iso}$, plotted as $x$ symbols, are from flares associated with a short burst.}
\label{fig:Epeak_Eiso_gold}
\end{figure}

In Figure \ref{fig:Epeak_Eiso_gold}, $E_{iso}$ (obtained as above using Band function spectral fits) is shown as a function of the redshift corrected $E_{peak}$ for only {\it{Gold}} flares that have a redshift measurement.  Due to the paucity of jet break measurements, we can not calculate E$_{\gamma}$, which corrects E$_{iso}$ by accounting for the jet opening angle, thus we can not explore the tighter E$_{peak}$--E$_{\gamma}$ relationship reported for GRB prompt emission by \citet{ghi05}.  

While it does seem clear that the flares involve a peak energy that is significantly lower than the more typical hundreds of keV observed for the initial GRB prompt emission, it is not clear if there is a strong relationship of this peak energy with E$_{iso}$, due to the large error bars and the limited sample.  The relatively low E$_{peak}$ in the X-ray band is, of course, expected since the flares are observed as increases in the X-ray band, which are often not accompanied by measurable increases in other bands.  Unfortunately, due to the size of the error bars on E$_{peak}$, it is not at all clear whether there is a relationship for the flares that is similar to the E$_{peak}$--E$_{iso}$ correlation found for the prompt GRB emission.  The intriguing hint of a relationship evident in Figure \ref{fig:Epeak_Eiso_gold} will be explored in the future using more flares and broader spectral coverage. 

\section{Redshift Distribution}
\begin{figure}[h]
\includegraphics[scale=0.4]{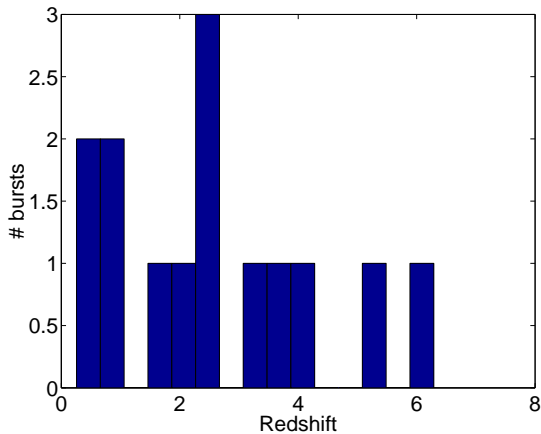}
\caption{Redshift distribution of GRBs with flares.}
\label{fig:redshift_distribution}
\end{figure}

This sample contains 14 GRBs that have a measured redshift.  The redshift distribution is shown in Figure \ref{fig:redshift_distribution}.  The mean redshift for these 14 GRBs is z=2.6.  This is consistent with the mean redshift of all {\it{Swift}} GRBs, which is between 2.5 and 2.8 \citep{bur06,jak06}.  This shows that the flaring GRBs are not drawn from a significantly different redshift distribution than the overall sample of GRBs. 

\section{Discussion \& Conclusions}

Based on this sample drawn from the first 110 {\it{Swift}} GRBs, it is clear that significant X-ray flares are produced frequently and at late times.  In this paper we have presented a detailed spectral analysis of 77 X-ray flares drawn from 33 GRBs.  Some of these GRBs had many flares (we find 7 significant flares in two of these GRBs, but there are probably even more temporally unresolved flares leading to observed intra-flare variability), while many GRBs had only one or two flares.  Each of the flares was treated as an individual event, and properties of the entire sample of flares were presented.  Although several studies of individual flares have been published recently, this paper (along with Paper I) provides a systematic study of a large sample of flares.  Furthermore, this paper is the first to correct for the effects of the underlying afterglow on the spectra of the flares.  For the flares in which the fluence of the underlying lightcurve is a significant fraction of the fluence of the flare, the effect of the underlying photons on the flare spectra is significant.

Several spectral models were fit to each flare.  These models included a simple absorbed power law, similar to the simple absorbed power law that typically fits the underlying afterglow lightcurve, as well as more complex models that are more akin to the spectral models that frequently provide a better description of GRB prompt emission, such as the GRB Band function \citep{ban93}.  For some flares, both simple and complex models provided a reasonable fit, while some flares had an improved fit by using a more complex model such as a Band function.  It is unlikely that the complete distribution of spectra was drawn from a pure power law distribution.  Spectra with curvature, such as the Band function, could be related to the instantaneous source spectrum, but it should be noted that it could also be caused by temporal evolution of the spectrum during the flare (this work averaged the spectrum over the entire flare) since some flares have shown spectral evolution in time \citep{god07,kri07}.  In any case, this result is similar to the results found for prompt emission from GRBs \citep{kan06,ban93}, in which power laws sometimes provided a reasonable fit to prompt emission while Band functions provided a better fit to the overall sample.  

It was also found that the photon spectral indices of the flare spectrum did not always match those of the underlying spectrum and the distributions were different from one another.  In those cases, this indicates a different population or mechanism for the production of the flare photons and the underlying afterglow lightcurve photons.  This result provides further evidence that flares result from some form of internal engine activity, particularly when evaluated in conjunction with the small ${\Delta}t/t$ values and further temporal analysis reported in Paper I, along with previous studies of individual bursts \citep{bur07,chi06,fal06}.  Within the context of the standard model, this would most likely involve very late internal shocks.  These internal shocks could, in principle, arise from a distribution of Lorentz factors for the shells from earlier internal engine activity, but this seems unlikely due to the inefficiency of the kinetic energy conversion that would result from these weak internal shocks \citep{zha06b,laz07}.  The late internal shocks are probably the result of late internal engine activity.

This sample contained 14 GRBs with a measured redshift, and the average redshift did not differ from the average redshift for all {\it{Swift}} GRBs, including those without flares.  This implies that late flares cannot be explained merely as redshifted multi-peaked emission from the initial prompt GRB.  This result is also supported by individual burst analyses that are corrected for redshift, such as that of \citet{cus06} in which a high redshift burst has very late flares in the burst rest frame.  All of this implies that the large fluence values for flares reported in this paper (sometimes comparable to the prompt GRB emission, and typically $\sim$10$\times$ less than prompt GRB) must be produced at very late times with peak energies in the X-ray band.  GRB progenitor models must be capable of producing this emission within a comparable energy budget to that which was previously applied to only the prompt GRB emission.  The fallback of material onto the central black hole after a stellar collapse could last for long time periods \citep{woo93, mac01} and lead to late internal engine activity, but the reduced luminosity of this model at late times means that it can not explain all flares.  Several models for continued activity of the central engine have been proposed (e.g. \citet{per06,fan06,dai06,kin05,pro05,kat97}).  These models, and others, must be evaluated within the context of the energy budget and the spectral parameters presented here.

This work has also attempted to explore the relationship between $E_{peak}$ and $E_{iso}$ for flares, in an effort to see if the Amati relationship \citep{ama06,ama02} is also present for flare events.  This is necessarily restricted to flares for which the GRB redshifts are known and the spectra have enough counts to constrain the parameters.  Unfortunately, this leads to only 18 flares (3 of which are from a short burst), and the parameters are not particularly well constrained, as shown in Figure \ref{fig:Epeak_Eiso_gold}.  Although a relationship may exist and there is an intriguing hint of a correlation similar to that reported by \citet{ama06} for GRB prompt emission, it is impossible to come to any firm conclusion due to the limited sample and large error bars.  By looking at more flares and analyzing data from more instruments over a wider energy band (thus improving the $E_{peak}$ constraint), this aspect of the flare study will be revisited in the future.

\acknowledgments
This work is supported at Pennsylvania State University by NASA contract NAS5-00136 and at Observatorio Astronomico di Brera by funding from ASI under grant I/R/039/04.



\end{document}